\documentclass[aps,preprint,tightenlines,superscriptaddress,
amsfonts,amssymb,amsmath,byrevtex,showpacs,nofootinbib,11pt]{revtex4-1}

\usepackage[polish,english]{babel}
\usepackage[applemac]{inputenc}
\usepackage[T1]{fontenc}

\usepackage{latexsym}
\usepackage{graphicx}
\usepackage{geometry}
\usepackage{epstopdf}
\usepackage{subfig}
\usepackage{color}

\usepackage{xcolor}
\usepackage{pdfsync}
\usepackage{fancyhdr}
\usepackage{makeidx}
\usepackage[breaklinks,colorlinks,backref]{hyperref}
\hypersetup{
    colorlinks, 
    linktoc=all, 
    linkcolor=blue,  
  citecolor=hyptxt,
  urlcolor=blue}
  \hypersetup{
  citebordercolor=,
  filebordercolor=green,
  linkbordercolor=blue
}
\definecolor{hyptxt}{rgb}{0.7, 0.4, 0.9}
\usepackage{bibtopic}

\newcommand{\be}{\begin{equation}}
\newcommand{\ee}{\end{equation}}

\begin{document}

\title{Quantum harmonic oscillator, entanglement in the vacuum and its geometric interpretation}

\author{Grzegorz Plewa\footnote{Email: {\tt greq771@gmail.com}}}
\affiliation{National Centre for Nuclear Research, Poland}

\begin{abstract}
Inspired by ER=EPR conjecture we present a mathematical tool providing a link between quantum entanglement and the geometry of spacetime. We start with the idea of operators in extended Hilbert space which, by definition, has no positive definite scalar product. Adopting several simple postulates we show that a quantum harmonic oscillator can be constructed as a positive definite sector in that space. We discuss the two-dimensional oscillator constructed in such a way that the ground state is maximally entangled. Being a vector in the Hilbert space, it has also a non-trivial expansion in a bigger extended space. On one hand, the space is not free of negative norm states. On the other hand, it allows one to interpret the ground state geometrically in terms of $AdS_3$. The interpretation is based solely on the form of the expansion, revealing certain structures at the boundary and in the bulk of $AdS_3$. The former correspond to world lines of massless particles at the boundary. The latter resemble interacting closed strings.
\end{abstract}

\maketitle

\tableofcontents

\section{Introduction}

ER=EPR \cite{Maldacena:2013xja,Susskind:2014yaa,Susskind:2014moa,Susskind:2016jjb} conjecture provides an interesting link between quantum entanglement and gravity, relating connected black holes to quantum entanglement. This utilizes earlier observations presented in \cite{VanRaamsdonk:2009ar,VanRaamsdonk:2010pw} and suggesting that quantum entanglement may be the fundamental concept explaining spacetime. In particular, it was shown that connectivity of space is closely related to entanglement. Breaking it causes disconnection of regions in space. In a broader sense, the connection predicted by the ER=EPR hypothesis can be extended to an even larger class of maximally entangled states, including e.g. the vacuum of relativistic QFT. What is more, recently it was suggested \cite{Susskind:2017ney} that quantum mechanics and gravity are actually not separate subjects. Even if the gravitational field is weak, their presence can be interpreted as manifestation of some quantum effects and vice versa. The last observation emerges from combination of various aspects of ER=EPR, AdS/CFT and black hole physics. There is a one missing point, however. Namely, we still cannot say much about the nature of the connection. For instance, if a maximally entangled state can be interpreted geometrically, what is the exact form of the corresponding geometry and how can it be found?

Trying to answer this, we propose a novel derivation of a quantum harmonic oscillator. The idea is to associate states in a given Hilbert space with vectors in some bigger space, large enough to incorporate a link with geometry. Looking for a potential candidate we consider the concept  of an extended Hilbert space, i.e. a Hilbert space without a positive definite scalar product. The idea of extended Hilbert space is not new and has already been discussed (see e.g. \cite{PhysRev.123.2183,doi:10.1143/PTP.59.972}). One of the reasons behind this concept is that positive definite Hilbert space is incompatible with manifest covariance of gauge invariant theories. Extended Hilbert space is also a natural generalization from gravitational point of view. In particular, the presence of negative norm states can be intuitively justified by the fact timelike directions in spacetime are also represented by negative norm vectors. Allowing them to be present already from the beginning can be a way to incorporate geometry. Here it is worth underlying this does not necessarily mean physical states have negative norms. As usual we assume they are necessarily positive definite and, as such, belong to a positive definite sector in the extended Hilbert space. Being a part of it, however, they can be connected with even more geometric objects. Searching for the link and understanding the role of quantum entanglement in the connection will be the main motivation for all calculations presented in this paper.

We start the analysis considering certain types of operators in the extended Hilbert space. They will be chosen in such a way that the definitions allow one to reproduce quantum harmonic oscillator as nothing but a positive definite sector in the space. We show that physical states have a unique expansion in the extended Hilbert space. This will provide additional information about the quantum system. For a class of maximally entangled states the information reveals an interesting connection with spacetime. The latter will be found as an emergent concept, specific the structure of states.

This paper is organized as follows. In section \ref{extended} we introduce the formalism, defining the operators and finding the algebra. We identify the Hilbert space of quantum harmonic oscillator as a positive definite sector in a more general extended Hilbert space. Considering the two-dimensional oscillator we reformulate it in such a way that the ground state is maximally entangled. We show that breaking the entanglement in the vacuum leads to creation a quantum of energy in a process similar to the Hawking effect. In section \ref{geometric} we examine the internal structure of the maximally entangled vacuum, provided by its expansion in the extended space. We observe that the expansion is nothing but a superposition of another orthogonal maximally entangled states, representing points in $AdS_3$. We find the geometric interpretation of the vacuum in a form of two distinct geometrical structures. The first looks like worldlines of massless particles at the boundary. The second resemble interacting closed strings in the bulk. In section \ref{sum} we discuss and summarize the results.

\section{Extended space}
\label{extended}

Below we show that the Hilbert space of a quantum harmonic oscillator can be found at a quantum level as a subspace of a bigger Hilbert space without a positive definite scalar product. The strategy will be first adopting necessarily abstract postulates and then examining the consequences. 

\subsection{External operators}
We start with the following definitions
\newtheorem{def1}{Definition}
\begin{def1}
\label{defi1}
Let $F_0$ be an extended Hilbert space, i.e. the Hilbert space with no necessarily positive definite scalar product. Let $\{ e_i,\partial_i \}_{i=1}^{N}$ stands for a set of pairs of operators acting on states in $F_0$, and defined such that for any $|\phi_a\rangle \in F_0$, one has
\begin{align}
\label{zera}
\langle \phi_b | e_{i_1}^{n_1}...e_{i_k}^{n_k}  | \phi_a \rangle  = \langle \phi_b | \partial_{i_1}^{n_1}...\partial_{i_k}^{n_k}  | \phi_a \rangle:= 0,
\\[1ex]
\label{komutatory}
[e_i,e_j]:=0, \quad [\partial_i,\partial_j]:=0,
\\[1ex]
\label{diffeo1}
\partial_i \, e_{j_1}...e_{j_n} | \phi_a \rangle := \delta_{i j_1} e_{j_2}...e_{j_n} | \phi_a \rangle +...+ \delta_{i j_n} e_{j_1}...e_{j_{n-1}} | \phi_a \rangle,
\\[1ex]
\label{diffeo2}
e_i \, \partial_{j_1}...\partial_{j_n} | \phi_a \rangle := - \, \delta_{i j_1} \partial_{j_2}...\partial_{j_n} | \phi_a \rangle -...- \, \delta_{i j_n} \partial_{j_1}...\partial_{j_{n-1}} | \phi_a \rangle,
\\[1ex]
\label{eherm}
e_i^\dagger:=e_i,
\end{align}
for $N$, $i_k$,  $j_k$  $n_k \in \mathbb{N}$. We call $e_i$ and $\partial_j$ the external operators.
\end{def1}
The definition above is to be understood as follows. The extended Hilbert space stands for a Hilbert space being not free of negative norm states. As mentioned in the introduction, the main reason for considering something like this is the belief this could be an option to incorporate gravity. The next step of the construction is defining the operator content and specifying the structure we wish to discuss. This is done introducing operators of two types, denoted as $e_i$ and $\partial_j$. They were  called  {\it external} because if, in particular, $F_0$ stands for a positive definite Hilbert space, then they go outside the space resulting in zero norm states of the form $ e_1 e_2 | \phi_a\rangle$, $\partial_1 | \phi_a\rangle$, etc. This is a direct consequence of the rule \eqref{defi1}. However, eq. \eqref{zera} is more than to say the external operators generate zero norm states. All expectation values of operators of the same type vanish as well. The postulate \eqref{zera} was dictated by technical simplicity. Considering complicated expressions built out of external operators, it would be extremely convenient to make some of them vanishing identically; something which is guarantied adopting the condition \eqref{zera}.

What makes the definition \ref{defi1} non-trivial are the rules \eqref{diffeo1}-\eqref{diffeo2}, supplementing the construction with operations decreasing the number of external operators in a given state. In particular, they allow transformations of the zero norm states back into the original vectors in $F_0$. For instance, if $|v\rangle = e_1^2 | \phi\rangle$ then $\frac{1}{2} \partial_1^2|v\rangle = |\phi \rangle$. Introducing \eqref{diffeo1} we require each of the operators $\partial_i$ is defined as eliminating one $e_i$ from the state. The same holds for $e_i$, however, the transformation comes with the extra minus sign. Its origin can be justified by the following observation. Let $E[F_0]$ stands for a vector space over the complex numbers, spanned by zero norm states resulted in products of operators $e_i$ acting on states in $F_0$. Similarly, let $D[F_0]$ be the analogous space built out of $\partial_i$ operators. Note that all states in $E[F_0] \oplus D[F_0]$ space are orthogonal to states in $F_0$. In what follows $E[F_0] \oplus D[F_0]$ decouples from $F_0$. It can be checked that eqs. \eqref{diffeo1}-\eqref{diffeo2} imply the following relation
\be
\label{dekom}
[\partial_i, e_j] = \delta_{ij}.
\ee 
The last holds true restricting solely to $E[F_0] \oplus D[F_0]$. The later will be of special importance and we call it the {\it extended space}. The additional minis sign in \eqref{diffeo2} is dictated by an observation that the commutator $[\partial_i, e_j]$ would not specify well-defined algebra in the extended space if the sign is chosen incorrectly. In particular, instead of \eqref{diffeo2} one may consider a modified rule 
\be
e_i \, \partial_{j_1}...\partial_{j_n} | \phi_a \rangle := c \, \delta_{i j_1} \partial_{j_2}...\partial_{j_n} | \phi_a \rangle +...+c \, \delta_{i j_n} \partial_{j_1}...\partial_{j_{n-1}} | \phi_a \rangle,
\ee
where $c$ is some constant to be fixed. Calculating the commutator $[\partial_i, e_j]$ one finds the algebra in the extended space is closed  if and only if $c=-1$.

The reason why we are interested in the commutator \eqref{dekom} is the analogy to  position and momentum, the basic ingredients needed to construct the harmonic oscillator (in the sense that creation and annihilation operators are linear combinations of these operators). Indeed, rewriting \eqref{dekom} as $[e_i, -i \partial_j] = i \delta_{ij}$, one identifies $e_i$ as playing the role of the position, whereas $-i \partial_j$, momentum. This is also supported by the condition \eqref{eherm}. This is not to say $e_i$ are Hermitian operators in the standard sense. Instead, this is only a definition of Hermitian conjugate of the object $e_i$. Similarly, combining eqs. \eqref{diffeo2}-\eqref{eherm} gives
\be
\partial_i^\dagger = -\partial_i.
\ee
In fact, this is the only way of making the conjugate $\partial_i^\dagger$ consistent with the definition \eqref{defi1}. It can be shown that both $e_i$ and $-i \partial_i$ are indeed Hermitian operators in the sense that one can find explicitly the space of their eigenstates\footnote{More specifically, the operators are Hermitian if they are restricted to these spaces. The latter are non-trivial subspeces in the extended Hilbert space.}. However, we do not discuss this in the paper. We restrict only to the observation that external operators, being analogs of position and momenta, are potentially interesting elements of a more complicated construction.

Talking about the analogy we should keep in mind it is incomplete since $e_i$ and $\partial_i$ are not operators in the Hilbert space. Instead, they belong to a bigger extended space $F_0 \oplus E[F_0] \oplus D[F_0]$. We call it the {\it maximally extended space}. It differs from its subspace, the extended space $ E[F_0] \oplus D[F_0]$, only by $F_0$. It turns out, this is a huge difference. In particular, the relation \eqref{dekom} takes a more complicated form in the maximally extended space
\be
\label{gendekom}
[\partial_i, e_j] = \delta_{ij} \hat{\eta},
\ee 
where 
\begin{align}
\nonumber
\hat{\eta} | \phi \rangle &:= 2 | \phi \rangle \,:  \quad  |\phi \rangle  \in F_0,
\\[1ex]
\hat{\eta} | \phi \rangle &:= | \phi \rangle  \,: \quad  |\phi \rangle  \in E[F_0] \oplus D[F_0].
\end{align}
The fact the external operators satisfy the well known algebra in the extended, but not in the maximally extended space suggests that the former is of special importance. As we shall see, in practical applications $F_0$ is completely meaningless. The only thing what matters is the structure in the extended space.

\subsection{Creation and annihilation operators}

Despite external operators being defined as generating zero norm states, their superpositions can lead to normalized vectors. For instance, it is easy to check that $\frac{1}{\sqrt{2}}(e_i\pm \partial_i)|\phi\rangle$ are unit, respectively negative and positive norm states.  This reflects the fact that a combination of states built out of a pair of operators $(e_i,\partial_i)$ acting on a given vector $|\phi\rangle \in F_0$, specifies a pair of orthogonal vectors. Requiring they are unit, one finds them in the form $\alpha_{\theta_i}^\dagger|\phi \rangle$, $\alpha_{\theta_i}|\phi \rangle$, where
\begin{align}
\label{alphafinal}
\alpha_{\theta_i} := (\theta^1_i + i \theta^2_i) e_i + \left(\frac{1-2 \theta^2_i \theta^3_i}{2 \theta^1_i}+ i \theta^3_i  \right) \partial_i
\end{align}
and  $ (\theta_i^a) := ( \theta^1_i,\theta^2_i,\theta^3_i ) $ stand for real coefficients:\footnote{Do not confuse the label $i$ in $\theta^a_i$  with the imaginary number $i$ multiplying $\theta^3_i$ in \eqref{alphafinal}.}, $\theta^1_i \in \mathbb{R}\setminus\{0 \}$, $\theta^2_i,\theta^3_i \in \mathbb{R}$.  For the sake of simplicity, we adopt a simplified convention writing $\alpha_i:= \alpha_{\theta_i}$ in short. The form \eqref{alphafinal} was found searching for a linear combination of external operators specifying orthonormal vectors. Each pair $(e_i,\partial_i)$ acting on a vector $|\phi\rangle \in F_0$ specifies two normalized states,  $\alpha_i| \phi \rangle$ and $\alpha_i^\dagger| \phi \rangle$. Quite interestingly,  both they are given by a single operator $\alpha_i$ and its Hermitian conjugate. The corresponding states are orthonormal in the sense that
\begin{align}
\nonumber
&\langle \phi| \alpha_i  \, \alpha_j |\phi \rangle = \langle \phi| \alpha_i^\dagger  \, \alpha_j^\dagger  |\phi \rangle = 0,
\\[1ex]
&\langle \phi| \alpha_i \, \alpha_j^\dagger |\phi \rangle = - \langle \phi| \alpha_j^\dagger  \, \alpha_i  |\phi \rangle = \delta_{i j} \| \phi \|,
\end{align}
where
\be
\| \phi \| := \langle \phi | \phi \rangle.
\ee
Note that $\alpha_i | \phi \rangle$ are negative norm states. This is because the maximally extended space, as a extended Hilbert space, is not positive definite. In particular, there is no linear combination of external operators leading to two positive norm states. 

Operators \eqref{alphafinal} satisfy the following commutation relations:
\begin{align}
\label{crealgebra}
&[\alpha_i,\alpha_j^\dagger] = \delta_{i j} \hat{\eta},
\\[1ex]
\label{supcrealgebra}
&[\alpha_i,\alpha_j]=[\alpha_i^\dagger,\alpha_j^\dagger]=0.
\end{align}
Eqs. \eqref{crealgebra}-\eqref{supcrealgebra} were derived under the assumption the operators are taken at a single, fixed point in the parameter space $\theta_i^a$ (i.e. the real parameters $\theta_i^{1...3}$ are fixed). For different points, $\theta_i^a$, ${\theta'}_i^a$, one can always express $\alpha_{\theta_i'} $ in terms of $\alpha_{\theta_i}$, using the identity
\be
\label{alphatransition}
\alpha_{\theta_i'}   = - \| \phi \|^{-1} \langle \phi |  \alpha_{\theta_i}^\dagger \alpha_{\theta_i'}  | \phi \rangle \alpha_{\theta_i} + \| \phi \|^{-1} \langle \phi |  \alpha_{\theta_i}  \alpha_{\theta_i'}  | \phi \rangle \alpha_{\theta_i}^\dagger.
\ee
Restricting to the extended space, the commutation relations  \eqref{crealgebra}-\eqref{supcrealgebra} take a simple form of the standard algebra of creation and annihilation operators
 \begin{align}
\nonumber
&[\alpha_i,\alpha_j^\dagger] = \delta_{i j},
\\[1ex]
&[\alpha_i,\alpha_j]=[\alpha_i^\dagger,\alpha_j^\dagger]=0.
\label{alphasimp}
\end{align}
Again, the  analogy is incomplete since  $\alpha_i | \phi \rangle \neq 0$ in general, and the negative norm states are present. Nevertheless, somewhat abusing the terminology, we call $\alpha_i $, $\alpha_j^\dagger$ {\it ladder operators}. As in case of position and momentum, the relations \eqref{alphasimp} emphasize the importance of the extended space. It is worth to underline that the algebra \eqref{alphasimp}, or more general commutation relations \eqref{crealgebra}-\eqref{supcrealgebra}, were not postulated but are direct consequence of the form \eqref{alphafinal} and the rules given by definition \ref{defi1}. Also note that the ladder operators were found searching for the simplest normalized states built out of combination of external operators. Hence, despite that we have found the algebra \eqref{alphasimp} in the end, the procedure of finding the ladder operators differs significantly from the standard derivation of creation and annihilation operators.

We close the discussion observing that either in the form \eqref{alphasimp}, as well as in the initial definition \eqref{defi1}, the operators are ''discrete'' i.e. they are labeled by discrete indexes. Clearly, this is not the only possibility. A consistent generalization to the continuous case is presented in appendix \ref{genext}. However, due to technical simplicity, in the rest of this paper we restrict ourselves to the operators labeled by, at most, two discrete labels.

\subsection{Positive definite sectors}

We now continue examining the maximally extended space specified by pairs of external operators acting on states in the fixed extended Hilbert space $F_0$. Having found the algebra \eqref{alphasimp}, we identify positive definite sectors in the extended space looking for a candidate for a well-defined Hilbert space. The latter is expected to be closely related with the algebra.

For the sake of simplicity, consider pairs of operators $(e_i, \partial_i)$ acting on a single, fixed normalized vector $|\phi \rangle \in F_0$. We call it the {\it fiducial vector}. Note that a single pair acting on the fiducial vector specify infinitely-dimensional maximally extended space of the form $F_0 \oplus E[F_0] \oplus D[F_0]$, where $F_0$ is a one-dimensional space composed of a single fiducial vector. Now, suppose we wish to identify a positive definite sector in the extended space $E[F_0] \oplus D[F_0]$. Recalling the algebra \eqref{alphasimp}, the simplest and most natural way is to construct a state $|\theta^+_i \rangle \in E[F_0] \oplus D[F_0]$, defined so that it is annihilated by $\alpha_i$:
\be
\label{cdef1}
\alpha_i |\theta^+_i \rangle = 0.
\ee
As a vector in the extended space, $|\theta^+_i \rangle$ is orthogonal to the fiducial vector
\be
\label{cdef2}
\langle \phi | \theta^+_i \rangle = 0.
\ee
The algebra \eqref{alphasimp} guaranties that any state of the form $(\alpha_i^\dagger)^n |\theta^+_i \rangle$, $n \in \mathbb{N}$, has a positive norm if the same holds for $|\theta^+_i \rangle$ as well. Being able to find it we get a positive definite sector related directly to the algebra. Note that the state is labeled by the same index $i$ as the pair $(e_i,\partial_i)$. For different pairs we expect to get different states.

It turns out that eqs. \!\eqref{cdef1}-\eqref{cdef2} uniquely determine the vector $|\theta^+_i \rangle$. The solution has the form  $| \theta^+_i \rangle = \hat{\theta}^+_i|\phi \rangle$, where
\be
\label{thetaplus}
\hat{\theta}^+_i = b^+(\theta_i) \sum_{n=0}^\infty \kappa^{-1}(n) \left(  (-1)^{n} \beta^n_i e_i^{2n+1} + \beta^{-n-1}_i  \partial^{2n+1}_i \right)
\ee
and
\be
\label{betakappa}
\beta_i := \frac{2 (\theta^1_i)^2 + 2 i \theta^1_i \theta^2_i}{1-2 \theta^2_i \theta^3_i}, \quad \kappa(n):=\prod_{k=0}^n(2k+1) = \frac{(2n+1)!}{2^n n!}.
\ee
Here $b^+_i(\theta_i)$ stand for normalization coefficients. Letting\footnote{Do not confuse labels of $r_i$ and $\varphi_i$ with the imaginary number.} $\beta_i = r_i e^{i \varphi_i}$, one finds
\be
\label{tocnorm}
\|\theta^+_i \| = \langle \phi| (\theta^+_i)^\dagger \theta^+_i | \phi \rangle = - \frac{ | b^+(\theta_i)|^2}{r_i} \left( \frac{ \textmd{arcsinh}(e^{-i \varphi_i}) }{  \sqrt{1+e^{- 2 i \varphi_1}}   } + \frac{\textmd{arcsinh}(e^{i \varphi_i}) }{  \sqrt{1+e^{2 i \varphi_1}}   } \right) \| \phi \|,
\ee
where $(r_i,\varphi_i)$, are  functions of $\theta_i^a$. Recalling the exact form of $\beta_i$ given by eq. \eqref{betakappa}, it can be checked that $\|\theta^+_i \| \propto - \| \phi \|$. The last holds independently of the choice of $\theta_i^a$. Hence $|\theta^+_i\rangle$ is a positive norm state for the negative norm fiducial vector. Since the choice of the norm is arbitrary, from now on, we let $\| \phi \| =-1$. Applying to \eqref{tocnorm}, the normalization condition $\|\theta^+_i \|=1$ gives the following normalization coefficient
\be
\label{bi}
b_i^+(\theta_i) = \sqrt{r_i}  \left( \frac{  \textmd{arcsinh} (e^{-i \varphi_i}) }{  \sqrt{1+e^{- 2 i \varphi_i}}   } + \frac{ \textmd{arcsinh}(e^{i \varphi_i}) }{  \sqrt{1+e^{2 i \varphi_i}}   } \right)^{-1/2}.
\ee
In principle, we should write $|b_i^+(\theta_i)|$ on the left. However, without losing the generality, one can assume that $b_i^+(\theta_i)$ are real. Now, let ${\cal{H}}^+_i$ stand for positive definite sectors determined by the vectors $|\theta^+_i\rangle$. With the normalization coefficients \eqref{bi}, the corresponding orthonormal bases read $\{\frac{1}{\sqrt{n!}}(\alpha^\dagger_i)^n|\theta^+_i\rangle \}_{n=0}^\infty$. Clearly, each ${\cal{H}}^+_i$ is nothing but a Hilbert space of (one-dimensional) quantum harmonic oscillator, where the state $|\theta^+_i \rangle$ plays a role of the vacuum\footnote{In this paper we use the terms {\it ground state} and {\it vacuum} interchangeably.}. This is the ground state of the Hamiltonian 
\be
\label{Hinit0}
\hat{H_i} = H_{0i} (\alpha_i^\dagger \alpha_i+b_i),
\ee
where $H_{0i}>0$ and $b_i$ are real parameters. The form  \eqref{Hinit0} can be justified searching for a Hermitian operator built out of minimal number of ladder operators and reproducing the whole Hilbert space as a space of eigenstates. Each state in the orthonormal basis $\{\frac{1}{\sqrt{n!}}(\alpha^\dagger_i)^n|\theta^+_i\rangle \}_{n=0}^\infty$ is nothing but the eigenvector of the Hamiltonian \eqref{Hinit0}. Identifying $H_{0i}=\hbar \omega_i$ and $b_i=\frac{1}{2}$, we get exactly what is expected for the quantum harmonic oscillator. However, with the lack of classical system, we cannot say much about these parameters except the fact they can be fixed referring to standard results. Still, we keep calling the resulting positive definite sector the harmonic oscillator. This means we will interpret $|\theta^+_i\rangle$ as the vacuum, treating the operator \eqref{Hinit0} as the Hamiltonian.

At this point it is worth mentioning the additional difference compared to standard analysis. Namely, eq. \eqref{cdef1} imposes a non-trivial constraint in the extended space. Solving it one finds the vacuum as a superposition of infinite number of states, specified by products of external operators acting on the fiducial vector. Alternatively, the expansion can be expressed in terms of ladder operators $\alpha_i$, $\alpha_i^\dagger$. Regardless of the parametrization, the vacuum has a unique internal structure. Here the {\it internal} refers to the fact the structure is meaningless from perspective of the Hilbert space, but it is present in the extended space, provided by the form \eqref{thetaplus} of the vacuum. This is the place where an additional information can be stored. We will examine this in the next section, discussing geometrical interpretation of the vacuum.  

In addition to ${\cal{H}}^+_i$, there are another positive definite sectors specific the algebra \eqref{alphasimp}. Consider the following modified  version of the condition \eqref{cdef1}: 
\be
\label{cmdef1}
\alpha_i^\dagger |\theta^-_i \rangle = 0.
\ee
That is, we now require the states to be annihilated by $\alpha_i^\dagger$. Demanding they are orthogonal to the fiducial vector (as vectors in the extended space), one finds $| \theta^-_i \rangle = \hat{\theta}^-_i |\phi \rangle$, where
\be
\label{thetaminus}
\hat{\theta}^-_i  = b^{-}_i(\theta_i) \sum_{n=0}^\infty \kappa^{-1}(n) \left( {\beta_i^*}^n e_i^{2n+1} + (-1)^{n+1} {\beta_i^*}^{-n-1}  \partial_i^{2n+1} \right).
\ee
Here $b^{-}_i(\theta_i)$ are the corresponding normalization coefficients. One can check that 
\be
\label{bcondi}
|b_i^-(\theta_i)| = |b_i^+(\theta_i)|,
\ee
and
\be
\label{bnorm}
\| \theta^+_i \| = - \| \theta^-_i \|, \quad \langle \theta^-_i | \theta^+_i \rangle = 0.
\ee
Resulting in letting  $\| \phi \|=-1$ one concludes that $|\theta^-_i\rangle$ are negative norm states. Each of them specifies a positive definite sector ${\cal{H}}^-_i$, spanned by the basis  
\be
\{ \frac{1}{ \sqrt{(2n-1)}!} \alpha^{2n-1}_i | \theta^-_i \rangle \}_{n=1}^\infty.
\ee
In contrast to ${\cal{H}}^+_i$ the state $|\theta^-_i\rangle$ is not a part of the sector. The space ${\cal{H}}^-_i$ has also no clear physical meaning. It looks like a quantum harmonic oscillator for which only every second excited state is taken into account. An interesting question is whether the two sectors ${\cal{H}}^\pm_i$ overlap or are independent. In the latter case, one could consider a bigger space ${\cal{H}}^-_i \oplus {\cal{H}}^+_i$, asking if this is a well-defined Hilbert space. It can be verified that a general superposition of states belonging to the two corresponding spaces, i.e. ${\cal{H}}^-_i$ and ${\cal{H}}^+_i$, is not a well-defined vector in the Hilbert space because the scalar product diverges. For instance, for $|v\rangle = c_1 \alpha_i^\dagger | \theta_i^+ \rangle + c_2 \alpha_i | \theta_i^- \rangle$, $c_1,c_2 \in \mathbb{C}$, one finds\footnote{For a suitable choice of $c_1$, $c_2$, however, the norm can be finite.} $\| v\|=-\infty$, i.e. the state is non-normalized. Hence, ${\cal{H}}^+_i$ and ${\cal{H}}^-_i$ should be regarded as separate Hilbert spaces. As we shall see, the problem can be partially circumvented by suitable redefinition of the scalar product. However, rather than searching for abstract spaces we will be interested  in reformulating the harmonic oscillator. This is why  we restrict solely to the sector ${\cal{H}}^+_i$. The only exception will be the vectors $| \theta_i^- \rangle$. Below we show they can be used providing an interesting generalization of the standard quantum harmonic oscillator, making the ground state entangled.

\subsection{Two-dimensional harmonic oscillator}

So far we discussed basic elements of the construction, finding the harmonic oscillator as a positive definite sector in the extended space. We now discuss a generalization involving a little bit more complicated, entangled states.

Restrict for simplicity to a two-dimensional case, considering two one-dimensional oscillators built out of two pairs of external operators denoted as  $(e_A,\partial_A)$ and $(e_B,\partial_B)$, acting on the two corresponding fiducial vectors, $| \phi \rangle_A$ and $| \phi \rangle_B$. Let  $\cal{H}$ stands for the Hilbert space of the resulting two-dimensional oscillator. The space is spanned by creation operators $\alpha_{A/B}^\dagger$ acting on the ground state in the form
\be
\label{vac}
|\theta\rangle := \hat{\theta}^+_A |\phi_A\rangle \otimes \hat{\theta}^+_B |\phi_B\rangle = | \theta_A^+ \rangle \otimes |\theta_B^+\rangle.
\ee 
Here $\hat{\theta}^+_i$ are given by eq. \!\eqref{thetaplus}, respectively for $i=A$ and $i=B$. The standard orthonormal basis reads  $\{ \frac{1}{\sqrt{n! m!}}(\alpha_A^\dagger)^n (\alpha_B^\dagger)^m |\theta\rangle  \}_{n,m=0}^\infty $, while the Hamiltonian is 
\be
\label{ham}
\hat{H} = H_0 \left( \sum_{i \in \{A,B\}} \alpha_i^\dagger \alpha_i + b  \right),
\ee
for $b \in \mathbb{R}$, $H_0>0$. Comparing with the standard results one identifies $H_0= \hbar \omega$ and $b=1/2$. However, here one should keep in mind the ladder operators were built out of external operators, i.e the objects which do not belong to the positive definite Hilbert space. At this point one may wonder what is the sense of making such analysis. As we recall, the main motivation is searching for the possible connection between geometry and entanglement. This will be the subject of separate analysis, presented in the next section. Below we construct more interesting class of maximally entangled states, which will be later interpreted geometrically. Looking for the potential candidate  we consider a reformulation of the two-dimensional oscillator such that the ground state is maximally entangled. 

One of the reasons for doing something like this is that the vacuum state of relativistic QFT is maximally entangled \cite{Harlow:2014yka}. Of course it is true that the two-dimensional oscillator has a little in common with quantum field theory, involving  uncountable number of oscillators. However, making the ground state entangled may provide a simple toy model, which reflects at the most superficial level what is known for vacuum QFT. At this moment we do not discuss more complicated and realistic states. Trying to identify geometric interpretation procedure, we start examining a very simple system. In fact, this is the reason why we consider the oscillator with the entangled ground state.

Having said that, we are now ready to make the desired generalization of the two-dimensional oscillator. To this end, consider the following vectors
\be
\label{zBell}
|\theta\rangle_\pm := \frac{1}{\sqrt{2}}\left( |\theta^+_A \rangle \otimes  |\theta^+_B \rangle \pm  |\theta^-_A \rangle \otimes | \theta^-_B \rangle \right),
\ee
where $|\theta_i^\pm\rangle = \hat{\theta}^\pm_i| \phi \rangle_i$ and $\hat{\theta}^\pm_i$ are given by eqs. \eqref{thetaplus} and \eqref{thetaminus}. As we recall, $| \theta^-_{A/B} \rangle$ are normalized negative norm states orthogonal to $| \theta^+_{A/B} \rangle$ (see eq. \eqref{bnorm}), which were found searching for another positive definite sectors in the extended space. In case of the vectors \eqref{zBell} they were added ''by hand'' to construct two maximally entangled Bell states. The entanglement consists in the fact both reduced density matrices are proportional to identity\footnote{This refers to the bases $\{ |\theta^+_A\rangle,|\theta^-_A\rangle \}$ and $\{ |\theta^+_B\rangle,|\theta^-_B\rangle \}$.}, i.e. $\rho_{A/B} = \frac{1}{2} id_{2 \times 2}$. The corresponding density operators read
\be
\label{densmat}
\hat{\rho}_i = \frac{1}{2} |\theta^+_i\rangle \langle \theta^+_i | + \frac{1}{2} |\theta^-_i\rangle \langle \theta^-_i |.
\ee
Note that despite negative norm states are present in \eqref{zBell}, the reduced density matrices (resulting from breaking the entanglement) are positive definite. Having postulated the entangled states \eqref{zBell} we have to clarify what does it mean in the context of the harmonic oscillator. In order to do so  observe that the abstract vectors $| \theta^-_{A/B}\rangle$ were constructed as annihilated by $\alpha_{A/B}^\dagger$. This leads to the identity
\be
\label{excident}
(\alpha_A^\dagger)^n (\alpha_B^\dagger)^m |\theta\rangle = \sqrt{2} (\alpha_A^\dagger)^n (\alpha_B^\dagger)^m |\theta \rangle_\pm.
\ee
Hence, each of the states $|\theta \rangle_\pm$ looks like a vacuum for the excited states. Any of them is not the true ground state, not even a vector in the Hilbert space.  Instead, the ''true'' vacuum is a superposition of the two states, i.e.
\be
\label{vacdec}
| \theta \rangle = \frac{1}{\sqrt{2}} \left( |\theta\rangle_+ + |\theta\rangle_-  \right).
\ee
Following the formal analogy behind eq. \eqref{excident} we would like to reformulate the Hilbert space in such a way that each of the abstract states \eqref{zBell} is a ground state of the quantum harmonic oscillator. The problem is that incorporating the states \eqref{zBell} the scalar product would be not well-defined. To be more specific, if $|\psi_{nm}\rangle$ stands for one of excited states of the oscillator, then  depending on the number of modes, the product $|_{\pm}\langle \theta|\psi_{nm}\rangle|$ either equals zero or diverges. This pathological behaviour forbids enlarging the Hilbert space by the abstract vectors \eqref{zBell}.

Fortunately, the problem  can be easily circumvented adopting small redefinition of the scalar product. Consider the following regularized version
\be
\label{regprod}
\langle \psi_1 | \psi_2\rangle_{reg}:= \left\{
\begin{array}{c l}	
     \langle \psi_1 | \psi_2\rangle: &  |\langle \psi_1 | \psi_2\rangle| < \infty\\
     0: & |\langle \psi_1 | \psi_2\rangle| = \infty.
\end{array}\right. 
\ee
It is clear that any excited state $|\psi_{nm}\rangle$ is now orthogonal to states \eqref{zBell}, i.e. $_{\pm}\langle \theta|\psi_{nm}\rangle = 0$. What is more, we can construct a new Hilbert space in such a way that either $|\theta\rangle_+$ or $|\theta\rangle_-$ will play a role of the ''vacuum''. In such reformulation all the excited states are given by the right hand side of eq. \eqref{excident}. The resulting Hilbert space will be trivially isomorphic with the Hilbert space of the standard two-dimensional oscillator.

The only missing thing is the Hamiltonian. As we recall, the Hamiltonian \eqref{ham} was found looking for a Hermitian operator involving minimal number of  ladder operators and reproducing the whole Hilbert space. We now expect to recognize at least one of the states \eqref{zBell} as the ground state of the Hamiltonian. The problem is that none of them is actually the eigenstate of operator \eqref{ham}. On the other hand, making small redefinition of eq. \eqref{ham}, the problem can be easily resolved. To be more specific, consider
\be
\label{hamprim}
\hat{H}' = H_0 \left( \sum_{i \in \{A,B\}} \alpha_i^\dagger \alpha_i + b  + 2 b |\theta^-_A \rangle  \langle \theta^-_A| \otimes |\theta^-_B \rangle  \langle \theta^-_B|  \right),
\ee
where we have added an additional projector $2 b |\theta^-_A \rangle  \langle \theta^-_A| \otimes |\theta^-_B \rangle  \langle \theta^-_B|$. This is a non-canonical term guarantying that both $|\theta\rangle_\pm$ are  eigenstates of the operator $H'$. Identifying $H_0 = \hbar \omega$, $b=\frac{1}{2}$ the operator \eqref{hamprim} reproduces the standard spectrum of the Hamiltonian, fixing the energy of the ground state to be equal $E_0 = \frac{1}{2} \hbar \omega$. The difference is that the two orthogonal states $|\theta\rangle_\pm$ correspond to the same energy level. The same holds for the superposition $c_1 |\theta\rangle_+ + c_2  |\theta\rangle_-$. 
Therefore, constructing the Hilbert space we have to choose the vacuum. In particular, letting $c_1=c_2=1$ we make the state separable, which corresponds to the original ground state \eqref{vac}. For $c_1=1,c_2=0$ or $c_1=0,c_2=1$ the ground state is  maximally entangled.  Whichever choice is made, we get the same spectrum of the Hamiltonian \eqref{hamprim}, finding the Hilbert space to be isomorphic with the Hilbert space of the two-dimensional harmonic oscillator. 

The Hamiltonian \eqref{hamprim} can be rewritten in a bit more elegant and familiar form 
\be
\label{hamprim2}
\hat{H}' = \hbar \omega \left( \sum_{i \in \{A,B\}} :\alpha_i^\dagger \alpha_i: + \frac{1}{2}  \right).
\ee
Here we have incorporated  $H_0 = \hbar \omega$, $b=\frac{1}{2}$, defining an order $::$ such that the ladder operators should be anti-normal ordered when acting on $|\theta^-_i \rangle$ ($\alpha_i$ stands to the left) and normal ordered ($\alpha_i$ stands to the right) when acting on $|\theta^+_i \rangle$ or any excited state. It is easy to check the operation effectively eliminates the need of the non-canonical term in the Hamiltonian.

Hawing said that, we have all elements needed for the reformulation. As we recall, the idea is to construct the oscillator in such a way the ground state will be maximally entangled. To this end, consider two Hilbert spaces denoted as  $\cal{H}'_\pm$ and defined as spanned by the following orthonormal  bases: $\{ |\theta\rangle_\pm, \frac{\sqrt{2}}{\sqrt{n! m!}}(\alpha_A^\dagger)^n (\alpha_B^\dagger)^m |\theta\rangle_\pm  \}_{n+m \geq 1}^\infty $. All of these vectors are eigenstates of the Hamiltonian \eqref{hamprim2}. Constructing Hilbert spaces we adopted regularized version of the scalar product \eqref{regprod}. Each of the two spaces $\cal{H}'_\pm$ is isomorphic with the Hilbert space of the two-dimensional harmonic oscillator. Since all the states of $\cal{H}'_+$ are orthogonal to $\cal{H}'_-$, the spaces $\cal{H}'_\pm$ can be regarded as subspaces of an even bigger Hilbert space, $\cal{H}'_+ \oplus \cal{H}'_-$.  Below we restrict to one of the corresponding sectors, either $\cal{H}'_-$ or $\cal{H}'_+$, writing $\cal{H}'_\pm$ in short.

Making the vacuum entangled one may ask about the potential physical meaning. Since the construction was motivated by what is known for the ground state of relativistic QFT, we should be ready to ask about possible physical consequences. In particular, an interesting question is what is the cost of breaking the entanglement in the vacuum. This results in creation of two maximally mixed states, described by density operators \eqref{densmat}. The corresponding average energies are
\be
\label{emix}
<E_{i}> = tr (\hat{H}'_i \hat{\rho}_i) = \hbar \omega, 
\ee
where $\hat{H}_i$ stands for the Hamiltonian of a one dimensional oscillator:
\be
\label{hamone}
\hat{H}'_i = \hbar \omega \left(  :\alpha_i^\dagger \alpha_i: + \frac{1}{2}  \right).
\ee
These were found under the assumption that quantum systems represented by the Hamiltonians $\hat{H}'_i$ are independent and non-interacting. Note that eq. \eqref{emix} is nothing but emission of a quantum of energy. Actually there are two of them, each for the two corresponding density matrices $\rho_{A/B}$. Each of them is twice as much as the energy of the initial maximally entangled ground state, $\frac{1}{2} \hbar \omega$. Therefore the emission process requires an additional energy source.

The fact that a single quantum can be extracted from the vacuum by breaking the entanglement can serve as qualitative illustration of Hawking process. Indeed, an intuitive explanation of the effect involves a pair of entangled virtual photons created just above the horizon of a black hole. One of them is absorbed by the hole, whereas the other escapes to infinity, becoming a quantum of Hawking radiation. Despite we have said nothing about black holes, the way of how eq. \eqref{hamone} was derived mimics the Hawking process. In particular, treating the maximally mixed states \eqref{densmat} as describing two non-interacting particles extracted from the vacuum \eqref{zBell} is a direct analogue of the Hawking emission. What is important here is that the process requires breaking the entanglement and the additional energy source. In case of the real emission one of the mixed states need to be absorbed by the hole to compensate the energy of the outgoing quantum. 

The analogy can be made even deeper finding entangled vectors \eqref{zBell} to be related with the state \eqref{vac} via a Bogoliubov transformation of creation and annihilation operators. Keeping in mind the formal analogy to Hawking process, it is also expected associating the transformation  with the presence of an accelerated observer. In fact, this the case for the Unruh effect: an accelerated observer traveling through the vacuum sees a thermal spectrum of particle excitations \cite{Harlow:2014yka}. By equivalence principle, the Unruh effect translates into the Hawking emission. The latter can be interpreted as a sort of the Unruh effect, caused by gravitational field of a black hole.

Despite the formal analogy, finding the Bogoliubov transformation connecting the states \eqref{zBell} and \eqref{vac} is technically complicated because of highly non-trivial form of $\hat{\theta}_i^\pm$ operators. This is one of the reasons why the states \eqref{zBell} were introduced ''by hand'' rather than being derived. From a physical point of view, however, it is expected such a transformation exists.

\section{Geometric interpretation of vacuum entanglement}
\label{geometric}

In the last section we reformulated the two-dimensional quantum harmonic oscillator in such a way, the ground state is maximally entangled and takes a simple form of one of the two Bell states \eqref{zBell}. We showed that breaking the entangled leads to the emission process resulting in extracting a quantum of radiation from the vacuum. In this section we look closer the states \eqref{zBell} themselves, examining their internal structure in the extended space and interpreting them geometrically in terms of anti-de Sitter space. 

\subsection{The internal structure}

We start considering two copies of the states \eqref{zBell}, $|\theta\rangle_\pm$ and $|\theta'\rangle_\pm$, labeled by two different points\footnote{Remember that ladder operators labeled by different parameters are actually not independent, but related by the transformation rule  \eqref{alphatransition}. From this perspective, taking into account different points in the parameter space seems to make no sense. However, it will be instructive to ignore this for a while, considering states built out of operators labeled by different values of these parameters.} $\theta^a_i$,  ${\theta'}^a_i$; $i=\{A,B\}$, in the parameter space. Consider the scalar product $_{\pm}\langle \theta'|\theta\rangle_\pm$. It can be verified that  $_{\pm}\langle \theta'|\theta\rangle _\pm = {_{\pm}\langle \theta|\theta'\rangle_\pm}$. This means the product is real. The last is a direct consequence of a symmetry behind the entanglement: making the states \eqref{zBell} maximally entanglement imposes specific constraints for the product. What is more, various expectation values of operators  in the product can be combined into four functional coefficients. More precisely,
\be
\label{vform1}
_{\pm}\langle \theta'|\theta\rangle _\pm = \sum_{n=0}^\infty \sum_{m=0}^\infty \left( \gamma^\pm_{n m}(\theta') \delta^\pm_{n m}(\theta)+ \gamma^\pm_{n m}(\theta) \delta^\pm_{n m}(\theta') +  \zeta^\pm_{n m}(\theta') \xi^\pm_{n m}(\theta)+\zeta^\pm_{n m}(\theta) \xi^\pm_{n m}(\theta') \right).
\ee 
Here $\gamma^\pm_{nm}$, $\delta^\pm_{nm}$, $\zeta^\pm_{nm}$, $\xi^\pm_{n m}$ are real functional coefficients given explicitly in appendix \ref{coef1}. Defining
\begin{align}
\nonumber
\chi^{\pm 0}_{nm}(\theta):=\frac{1}{\sqrt{2}}(\delta^\pm_{nm}(\theta)-\gamma^\pm_{nm}(\theta)), \quad \chi^{\pm 1}_{nm}(\theta):=\frac{1}{\sqrt{2}}(\zeta^\pm_{nm}(\theta)+\xi^\pm_{nm}(\theta)),
\\[1ex]
\label{mapa1}
\chi^{\pm 2}_{nm}(\theta):=\frac{1}{\sqrt{2}}(\gamma^\pm_{nm}(\theta)+\delta^\pm_{nm}(\theta)), \quad \chi^{\pm 3}_{nm}(\theta):=\frac{1}{\sqrt{2}}(\zeta^\pm_{nm}(\theta)-\xi^\pm_{nm}(\theta)),
\end{align}
one rewrites \eqref{vform1} as
\begin{align}
\nonumber
_{\pm}\langle \theta'|\theta\rangle _\pm = \sum_{n=0}^\infty \sum_{m=0}^\infty \Big( &- \chi^{\pm 0}_{nm}(\theta) \chi^{\pm 0}_{nm}(\theta')+\chi^{\pm 1}_{nm}(\theta) \chi^{\pm 1}_{nm}(\theta')+
\\[1ex]
\label{vform2}
&+\chi^{\pm 2}_{nm}(\theta) \chi^{\pm 2}_{nm}(\theta')-\chi^{\pm 3}_{nm}(\theta) \chi^{\pm 3}_{nm}(\theta')  \Big).
\end{align}
In particular, letting  ${\theta'}^a_i = \theta^a_i$, one finds
\be
\label{constr}
 _{\pm}\langle \theta|\theta\rangle _\pm =\sum_{n=0}^\infty \sum_{m=0}^\infty \Big( -(\chi^{\pm 0}_{nm}(\theta))^2 + (\chi^{\pm 1}_{nm}(\theta))^2 + (\chi^{\pm 2}_{nm}(\theta))^2 - (\chi^{\pm 3}_{nm}(\theta))^2 \Big) = 1,
\ee
where the last equality reflects the normalization constraint (\eqref{zBell} are unit vectors). It is worth to underline the factorization of the product \eqref{vform2} (or \eqref{vform1}) is a consequence of quantum entanglement of the state. In particular, if instead of \eqref{zBell}, one considers separate or not maximally entangled states, the product does not factorizes.

From a geometrical point of view the scalar product \eqref{vform1} has two interesting features we will utilize in a moment. The first are mentioned consequences of entanglement: the fact the scalar product is real and factorizes. The second consists in the form \eqref{vform2}. Note that any term in the sum with fixed $n$ and $m$ looks like a scalar product of two vectors in $\mathbb{R}^{2,2}$, defined as follows: $\vec{\chi}^\pm_{nm} = (\chi_{nm}^{\pm 0}(\theta),...,\chi_{nm}^{\pm 0}(\theta))$ and ${\vec{\chi}_{nm}'^\pm} = (\chi_{nm}^{\pm 0}(\theta'),...,\chi_{nm}^{\pm 0}(\theta'))$. The identification relies upon the form of functional coefficients. However, since the product \eqref{vform2} consists infinite number of such terms, each of them for different values $n$, $m$, it should be identified with a point in infinite-dimensional pseudo-hyperbolic space \cite{BADITOIU:2002nq}. The functional coefficients $\chi^{\pm \mu}_{nm}(\theta)$ are identified with coordinates in $\mathbb{R}^{\infty,\infty}$, while the normalization \eqref{constr} plays a role of the constraint determining the pseudo-hyperbolic space.

Fortunately, the interpretation above is not the only possible one, and one can identify even more interesting class of states. To see this, it is instructive to look closer to the structure of the states $|\theta\rangle_\pm$. It turns out, the latter can be recast as the following superposition
\be
\label{vac0}
|\theta\rangle_\pm = \sum_{n=0}^\infty \sum_{m=0}^\infty c_{nm} | \theta_{n m}\rangle_\pm,
\ee
where
\be
\label{svac}
| \theta_{n m}\rangle_\pm = \frac{1}{\sqrt{2}} N_{An} N_{Bm} \left( \hat{\theta}_{An}^+|\phi\rangle_A \otimes \hat{\theta}_{Bm}^+|\phi\rangle_B \pm \hat{\theta}_{An}^-|\phi\rangle_A \otimes \hat{\theta}_{Bm}^-|\phi\rangle_B  \right),
\ee
and 
\begin{align}
\label{splus}
&\hat{\theta}_{in}^+:= \kappa^{-1}(n)  \left( (-1)^n \beta_i^n e_i^{2n+1} + \beta_i^{-n-1} \partial_i^{2n+1} \right),
\\[1ex]
\label{sminus}
&\hat{\theta}_{in}^-:= \kappa^{-1}(n) \left( (\beta_i^*)^n e_i^{2n+1} + (-1)^{n+1} (\beta_i^*)^{-n-1} \partial_i^{2n+1} \right),
\\[1ex]
&c_{nm} = \tilde{c}_{An} \tilde{c}_{Bm},
\label{ctilde}
\\[1ex]
\label{Nc}
&\tilde{c}_{in} = b_i N_{in}^{-1}, \quad  N_{in}= \frac{\sqrt{(2n+1)!}\sqrt{r_i}}{ 2^{n+\frac{1}{4}}  n! [1+\cos((4n+2)\varphi_i)]},
\end{align}
$i = A,B$. Here $b_i=b_i^{+}(\theta_i)$ are given by eq. \eqref{bi}, while $\kappa(n)$ was defined in \eqref{betakappa}. As before we skipped the labels $\theta^a_{i}$ for convenience\footnote{For instance, $b_A =b_A(\theta_A)$, $r_B = r_B(\theta_B)$, etc.}. Note the form \eqref{svac} follows directly from eq. \eqref{zBell}. In fact, eq. \eqref{vac0} comes from  identification of the basic components in the superposition, i.e. the vectors $| \theta_{n m}\rangle_\pm$, and making them unit vectors. The last consists in adopting the additional normalization coefficients $N_{in}$ \eqref{Nc}, chosen such that the states \eqref{svac} are orthonormal:
\begin{align}
\label{snorm1}
|_{\pm}\langle \theta_{n' m'}|\theta_{n m} \rangle_\pm| &= \delta_{n' n}\delta_{m' m}, 
\\[1ex]
\label{snorm2}
_{\pm}\langle \theta_{n' m'}|\theta_{n m} \rangle_\mp &= 0.
\end{align}
(without  $N_{in}$ they would be only orthogonal). The absolute values on the left hand side of eq. \eqref{snorm1} are due to the fact that, depending on the choice of $n$, $m$, and the values of parameters $\theta^a_{i}$, some of the vectors \eqref{svac} have negative norms. More precisely,
\begin{align}
\label{cnem1}
\| \theta_{nm \pm} \| &= {_{\pm}\langle \theta_{n m}|\theta_{n m} \rangle_\pm} = l^2_{nm}(\theta),
\\[1ex]
\label{cnem2}
l^2_{nm}(\theta)&=(-1)^{n+m} \textmd{sign}[ {\cos((2n+1)\varphi_A)} {\cos((2m+1)\varphi_B)} ].
\end{align}
This shows that the negative norm states cannot be eliminated by suitable choice of the parameters $\theta^a_{i}$. Another observation is that normalizing the positive norm states to unity automatically guaranties the negative norm states are also unit vectors. From now on we call the abstract vectors \eqref{svac}  {\it partial states}. As we shall see in a moment, they span a convenient basis in the extended space.

In addition to be orthonormal (in the sense of \eqref{cnem1}), partial states \eqref{svac} can be interpreted as maximally entangled. As in case of the states \eqref{zBell} this lies in the fact that both the reduced density matrices are proportional to identity. The subtlety is that now we apply this criterion both to positive and negative norm states \eqref{svac}. The reason is that in both cases the reduced density matrices are positive definite. However, rather than treating the partial states as well-defined quantum mechanical objects, we will concentrate mostly on their geometrical aspects. In particular, we will interpret them geometrically, finding the entanglement to be essential for the interpretation.

Before doing this it is worth to look closer the amplitudes $c_{nm}$ in the superposition \eqref{vac0}. Since the negative norm states are present,  $\sum_{nm} |c_{nm}|^2 \neq 1$. Similarly, one cannot expect the coefficients\footnote{Constructing the states we assumed the normalization coefficients $b_i^\pm$ of $|\theta_i^\pm\rangle$ are real, which implies $c_{nm} \in \mathbb{R}$. However, a simple unitary transformation of the states $|\theta\rangle_\pm \rightarrow e^{i \phi}|\theta\rangle_\pm$ can make them complex. This is why despite in \eqref{vac0} they are formally real, we will treat them like complex numbers.} $|c_{nm}|^2 \leq 1$. Hence, $c_{nm}$ are not the standard probability amplitudes. On the other hand, one can check that the form \eqref{vac0} implies they are still constrained  in the sense that $|c_{nm}| \leq 4 \textmd{arcsinh}^{-1}(1) \approx 4.54$. The inequality is saturated  for $\varphi_{A/B}=n=m= 0$. The coefficients $c_{nm}$ play a role analogous to the standard probability amplitudes, measuring contribution of $nm$-th partial state to the superposition. In particular, if  $|c_{nm}|^2 \ll 1$ then the corresponding vector $|\theta_{nm}\rangle_\pm$ is meaningless from perspective of the scalar product. Taking this into account, we call $|c_{nm}|^2$ {\it effective contributions}\footnote{Possibly a better way is to consider renormalized effective contributions $\frac{1}{16}|c_{nm}|^2 \textmd{arcsinh}^{2}(1) \in [0,1]$. We intentionally ignore such details now.}.

Having said that, we can now continue discussing geometric interpretation of the partial states. By analogy to eq. \eqref{vform1} consider a product of two states of the form \eqref{svac}, labeled by two different set of parameters $\theta^a$ and ${\theta'}^a$. The product reads
\be
\label{scalar2}
{_{\pm}\langle \theta'_{nm}|\theta_{nm} \rangle_\pm} = \tilde{\gamma}^\pm_{nm}(\theta') \tilde{\delta}^\pm_{nm}(\theta) +\tilde{\gamma}^\pm_{nm}(\theta) \tilde{\delta}^\pm_{nm}(\theta') + 
\tilde{\zeta}^\pm_{nm}(\theta') \tilde{\xi}^\pm_{nm}(\theta) +\tilde{\zeta}^\pm_{nm}(\theta) \tilde{\xi}^\pm_{nm}(\theta'),
\ee
where  $\tilde{\gamma}^\pm_{nm}(\theta)$, $\tilde{\delta}^\pm_{nm}(\theta)$, $\tilde{\zeta}^\pm_{nm}(\theta)$, $\tilde{\xi}^\pm_{nm}(\theta)$ are real functional coefficients, given explicitly in appendix~\ref{coef1}. The scalar product above can be identified with a four-dimensional quadratic form acting on two vectors of the type $(\tilde{\gamma}^\pm_{nm}(\theta)$, $\tilde{\delta}^\pm_{nm}(\theta)$, $\tilde{\zeta}^\pm_{nm}(\theta)$, $\tilde{\xi}^\pm_{nm}(\theta))$ and $(\tilde{\gamma}^\pm_{nm}(\theta')$, $\tilde{\delta}^\pm_{nm}(\theta')$, $\tilde{\zeta}^\pm_{nm}(\theta')$, $\tilde{\xi}^\pm_{nm}(\theta'))$. Introducing new functional coefficients defined as
\begin{align}
\nonumber
X^{0}_{\pm nm}(\theta):=\frac{1}{\sqrt{2}}(\tilde{\delta}^\pm_{nm}(\theta)-\tilde{\gamma}^\pm_{nm}(\theta)), \quad X^{1}_{\pm nm}(\theta):=\frac{1}{\sqrt{2}}( \tilde{\zeta}^\pm_{nm}(\theta) + \tilde{\xi}^\pm_{nm}(\theta)),
\\[1ex]
\label{map0}
X^{2}_{\pm nm}(\theta):=\frac{1}{\sqrt{2}}(\tilde{\delta}^\pm_{nm}(\theta) +\tilde{\gamma}^\pm_{nm}(\theta) ), \quad X^{ 3}_{\pm nm}(\theta):=\frac{1}{\sqrt{2}}(\tilde{\zeta}^\pm_{nm}(\theta)-\tilde{\xi}^\pm_{nm}(\theta)),
\end{align}
the form can be easily diagonalized. The product reads
\begin{align}
\nonumber
\label{sosym}
{_\pm}\langle \theta'_{nm}|\theta_{nm} \rangle_\pm &= -X_{\pm nm}^{0}(\theta')X_{\pm nm}^{0}(\theta) + X_{\pm nm}^{ 1}(\theta')X_{\pm nm}^{1}(\theta) +
\\[1ex]
&+ X_{\pm nm}^{2}(\theta')X_{\pm nm}^{2}(\theta) - X_{\pm nm}^{3}(\theta')X_{\pm nm}^{3}(\theta).
\end{align}
Since $n$, $m$ are fixed, this is nothing but a product of two vectors $\vec{X}_{\pm nm}(\theta)$ and $\vec{X}_{\pm nm}(\theta')$ in $\mathbb{R}^{2,2}$, where $\vec{X}_{\pm nm}(\theta) := (X_{\pm nm}^{0}(\theta),...,X_{\pm nm}^{3}(\theta))$. 
Letting ${\theta'}^a_i = \theta^a_i$ and taking into account the normalization constraint \eqref{cnem1}, one finds they are unit
\be
\label{sosym2}
-(X^0_{\pm nm})^2 + (X^1_{\pm nm})^2 + (X^2_{\pm nm})^2 - (X^3_{\pm nm})^2 = l^2_{nm}.
\ee
Here $X^0_{\pm nm} = X^0_{\pm nm}(\theta)$ and $l^2_{nm}= l^2_{nm}(\theta) \in \{ -1,1\}$. Eq. \eqref{sosym2} can be interpreted as a normalization constraint for spacetime vectors. Depending on the values of parameters ($n$, $m$, $\varphi_{A/B}$), $\vec{X}_{\pm nm}(\theta)$ are either spacelike or timelike unit vectors in $\mathbb{R}^{2,2}$ (in either case there is a trivial isomorphism with unit vectors in $\mathbb{R}^{2,2}$). The scalar product has an internal symmetry of $SO(2,2)$ rotations. More precisely, any transformation of the  parameters $\theta^a$ resulting in $SO(2,2)$ rotation of $\vec{X}_{\pm nm}(\theta)$ is a symmetry of the product. 

The presence of $SO(2,2)$ symmetry and, in particular, the fact that the scalar product of two partial states takes the form of the scalar product of two unit vectors in  $\mathbb{R}^{2,2}$ has a nice geometrical illustration in terms of $AdS_3$ space. Indeed, the normalization constraint \eqref{sosym2} can be recognized as the standard condition used in constructing $AdS_3$ as embedding in $\mathbb{R}^{2,2}$  \cite{Bayona:2005nq}. Solving the constraint  \eqref{sosym2} one installs coordinates on the resulting three dimensional manifold. A convenient choice which will be adopted in this paper are Poincare coordinates \cite{Bayona:2005nq}. Since  $\mathbb{R}^{2,2}$ is four-dimensional, $AdS_3$ can be found in both cases $l^2_{nm} = \pm 1$. The only difference is that depending on the sign of $l^2_{nm}$, either $(X^0_{\pm nm},X^3_{\pm nm})$ or $(X^1_{\pm nm},X^2_{\pm nm})$ become timelike. Another important thing is  that the construction relies strongly upon the structure of the product \eqref{sosym}, not just the norm \eqref{sosym2} itself. Despite the latter is crucial identifying $AdS_3$ as embedding manifold in $\mathbb{R}^{2,2}$, the former is responsible for identifying real functional coefficients $X^\mu_{\pm nm}$ with coordinates in $\mathbb{R}^{2,2}$. Without this identification, the whole subsequent procedure would not make sense.

An important feature of the construction is that identifying the partial states with unit vectors in $\mathbb{R}^{2,2}$ we considered the product of two states labeled by two different parameters $\theta^a_i$ and ${\theta'}^a_i$. In fact, this is what lead us to $SO(2,2)$ group of internal symmetry and, finally, $AdS_3$ space. Here we should keep in mind that constructing the harmonic oscillator we treated $\theta^a_i$ as fixed parameters. Different values correspond to equivalent, but in different formulations of the Hilbert space. The difference lies in the fact the ladder operators are given by different  linear combinations of external operators  $e_i$ and $\partial_i$. Once a particular choice is made, it holds for the whole Hilbert space. In particular, constructing the two-dimensional oscillator in such a way that either  $|\theta\rangle_+$ or  $|\theta\rangle_-$ will be the ground state, one has to fix values of the parameters $\theta^a_i$.

On the other hand, for a fixed $\theta^a_i$ one could always consider a small perturbation in the parameter space  ${\theta'}^a_i = \theta^a_i + \delta \theta^a_i$ and then examine the product $_{\pm}\langle \theta + \delta \theta| \theta \rangle_\pm$. Making the deviation $\delta \theta^a_i$ to be small can be justified as identifying an additional (scalar) attribute associated to the state. This is not us unusual as one might think. For instance, all expansions in the extended space given by explicit form of the operators $\hat{\theta}^\pm_i$ are already the additional attributes assigned to vectors in the Hilbert space of quantum harmonic oscillator. The same holds for partial states as well. Involving a small perturbation  ${\theta'}^a_i = \theta^a_i + \delta \theta^a_i$ and considering the product $_{\pm}\langle \theta + \delta \theta| \theta \rangle_\pm$ is simply a way of extracting the additional information, hidden in the internal structure of the states. Since the parameters $\theta^a_i$ are fixed and since $ \delta \theta^a_i$ are infinitesimally small, instead of the whole anti de Sitter space, each partial state specifies a single point in $AdS_3$ (as we shall see in a moment, there are actually two of them). Below we look closer to the construction. 

Let $|\theta_{nm}\rangle_\pm$ be a fixed vector \eqref{svac}, chosen so that this is a negative norm state, e.g. $l^2_{nm}=-1$ ($n$, $m$ are fixed). Positive norm vectors will be discussed in a moment. Consider a small perturbation in the parameter space ${\theta'}^a_i = \theta^a_i + \delta \theta^a_i$ and the product ${_\pm}\langle \theta'_{nm}|\theta_{nm} \rangle_\pm$. According to eq. \eqref{sosym}, the four functional coefficients $X^\mu_{\pm nm}$ can be identified with vectors in $\mathbb{R}^{2,2}$. They are constrained by eq. \eqref{sosym2}. For the negative norm partial states ($l^2_{nm}=-1$) the constraint reads
\be
\label{sosym2minus}
-(X^0_{\pm nm})^2 + (X^1_{\pm nm})^2 + (X^2_{\pm nm})^2 - (X^3_{\pm nm})^2 = -1.
\ee
This can be solved introducing Poincare coordinates $(t_{\pm nm},x_{\pm nm},z_{\pm nm})$, defined by the standard formulas \cite{Bayona:2005nq}:
\begin{align}
\label{Pn1}
X^0_{\pm nm} &= \frac{-t^2_{\pm nm}+x^2_{\pm nm}+z^2_{\pm nm}+1}{2 z_{\pm nm}},
\\[1ex]
\label{Pn2}
X^1_{\pm nm} &= \frac{x_{\pm nm}}{z_{\pm nm}},
\\[1ex]
\label{Pn3}
X^2_{\pm nm} &= \frac{-t^2_{\pm nm} +x^2_{\pm nm} + z^2_{\pm nm}-1 }{2 z_{\pm nm}},
\\[1ex]
\label{Pn4}
X^3_{\pm nm} &= \frac{t_{\pm nm}}{z_{\pm nm}}.
\end{align}
These can be solved finding the Poincare coordinates
\be
\label{cdrminus}
t_{\pm nm} =(\pm) \frac{|X^3_{\pm nm}|}{|X^0_{\pm nm}-X^2_{\pm nm}|}, \quad x_{\pm nm} = \frac{X^1_{\pm nm}}{X^0_{\pm nm}-X^2_{\pm nm}}, \quad z_{\pm nm} = \frac{1}{X^0_{\pm nm}-X^2_{\pm nm}}. 
\ee
Here $(\pm)$ refers to the fact that there are two solutions representing opposite moments in time: $t_{\pm nm} = |X^3_{\pm nm}| |X^0_{\pm nm}-X^2_{\pm nm}|^{-1}$ and $t_{\pm nm} = -|X^3_{\pm nm}| |X^0_{\pm nm}-X^2_{\pm nm}|^{-1}$. In result each partial states, i.e. $|\theta_{nm}\rangle_+$ or $|\theta_{nm}\rangle_-$, can be associated with a single pair of points in $AdS_3$.

Now, suppose  $l^2_{nm}=1$, i.e. the partial states $|\theta_{nm}\rangle_\pm$ have positive norms. The constraint \eqref{sosym2} now takes the form
\be
\label{sosym2plus}
-(X^0_{\pm nm})^2 + (X^1_{\pm nm})^2 + (X^2_{\pm nm})^2 - (X^3_{\pm nm})^2 = 1.
\ee
Multiplying both sides of eq. \!\eqref{sosym2plus} by minus one, one can repeat the steps presented above and  install the Poincare coordinates. The only difference is that the former timelike coordinates $(X^0_{\pm nm},X^3_{\pm nm})$ now become spacelike, while the spacelike $(X^1_{\pm nm},X^2_{\pm nm})$ become timelike. The corresponding redefinition of the conditions \eqref{Pn1}-\eqref{Pn4} leads to the following identification
\be
\label{cdrplus}
t_{\pm nm} =(\pm) \frac{|X^2_{\pm nm}|}{|X^1_{\pm nm}-X^3_{\pm nm}|}, \quad x_{\pm nm} = \frac{X^0_{\pm nm}}{X^1_{\pm nm}-X^3_{\pm nm}}, \quad z_{\pm nm} = \frac{1}{X^1_{\pm nm}-X^3_{\pm nm}}.
\ee
Note that in both cases  $l^2_{nm}=\pm1$ we get the same result:  partial states are interpreted geometrically as pairs of points in $AdS_3$. It is worth underlying despite considering the perturbation ${\theta'}^a_i = \theta^a_i + \delta \theta^a_i$ is essential for geometric interpretation (because of the structure of the product \eqref{sosym}), it is meaningless from perspective of solving the constraint \eqref{sosym2} and installing the coordinates on the resulting manifold.

\subsection{Structures specific to the entangled ground state} 

Examining the structure of vectors \eqref{zBell} we isolated maximally entangled partial states, interpreting them as points in $AdS_3$. Now we come back to the original states \eqref{zBell}, asking about their potential geometrical interpretation. In contrast to the partial states the vectors \eqref{zBell} are of special importance  because each of them can be identified with the vacuum of a two-dimensional quantum harmonic oscillator. So far the states \eqref{svac} were interpreted geometrically, whereas \eqref{zBell} were not. It is natural to expect that having interpreted the basis vectors \eqref{svac}, we should be ready to say something about their superpositions \eqref{zBell}.

Since the states \eqref{zBell} are maximally entangled, according to the ER=EPR conjecture, they (at least) have a chance to be interpreted geometrically. Taking into account the fact they are superpositions of states already interpreted as points in $AdS_3$, it is natural to recognize them as specifying discrete regions in the space. Such interpretation would be a direct analogue of the standard construction of position eigenstates $|x\rangle$ in position representation. More precisely, the position eigenvectors are labeled by values of position in Euclidean space in such a way, the vectors are orthogonal, i.e.  $\langle x'|x\rangle = \delta^3(x'-x)$. Similarly, partial states are also orthogonal and translate into points in $AdS_3$. The main difference is that $|\theta\rangle_\pm$ describe discrete rather than continuous regions. 

Below we take a closer look the proposed geometrical interpretation of the states \eqref{zBell}. First of all, we should underline this is a secondary concept, based on the earlier geometric interpretation of the partial states and the ER=EPR. The latter is crucial, since it provides a basic criterion for the procedure: the states under consideration should be maximally entangled. In fact, this is the reason for adopting the basis built out of maximally entangled partial states \eqref{svac}. It seems it would make no sense asking about geometric interpretation of states which are not maximally entangled. In particular, the ground state \eqref{vac} is a superposition of two maximally entangled Bell states  \eqref{zBell}. However, since the state \eqref{vac} is separable, we do not expect it can be interpreted geometrically within the ER=EPR. This is also reflected by the construction. For instance, making the partial states {\it not} maximally entangled and repeating the procedure of taking the 
perturbation  $\theta^a_i \rightarrow \theta^a_i+\delta \theta^a_i$, one gets nothing interesting in result. The  scalar product does not factorize, it is not even real. This holds true for various different vectors in the extended space, including the separable state \eqref{vac}. It would be interesting to ask if there is a rigorous proof stating that the entanglement is indeed essential for geometric interpretation. If so, this would be an argument in favor of the ER=EPR conjecture.

Accepting the above, we are now ready to identify the regions specified by the structure of the states $|\theta\rangle_\pm$. Since they are superpositions of infinite number of partial states, their geometric representation involves infinitely many of discrete points in $AdS_3$, leading to a discrete structure in that space. For a dense enough number of points the latter can be approximated by continuous manifold. More specifically, in the Poincare coordinates the metric reads
\be
\label{AdS}
ds^2 = \frac{1}{z^2} \left( -dt^2+dx^2 + dz^2 \right).
\ee
Here we should keep in mind this is an approximation valid if the points are sufficiently close to each other. The goal will be now to examine this, finding structures specific the states \eqref{zBell} by plotting the corresponding points at a single graph. To this end we additionally impose a finite cut-off $\cal{N}$ for infinite superposition \eqref{vac0}.

\begin{figure}
\begin{minipage}{1\linewidth}
\centering
  \subfloat[]{\includegraphics[width=0.45\textwidth]{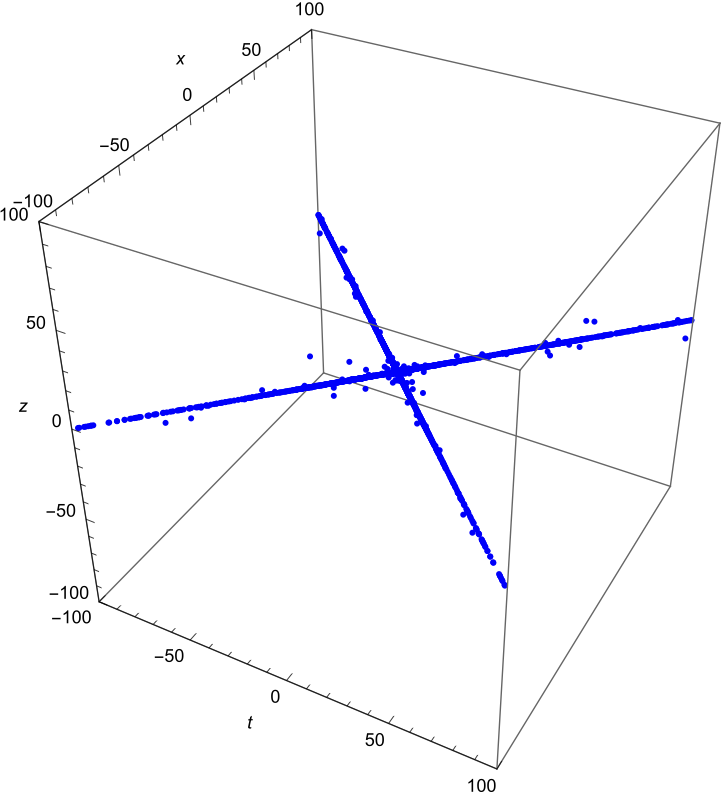}}
  \hfill
  \subfloat[]{\includegraphics[width=0.45\textwidth]{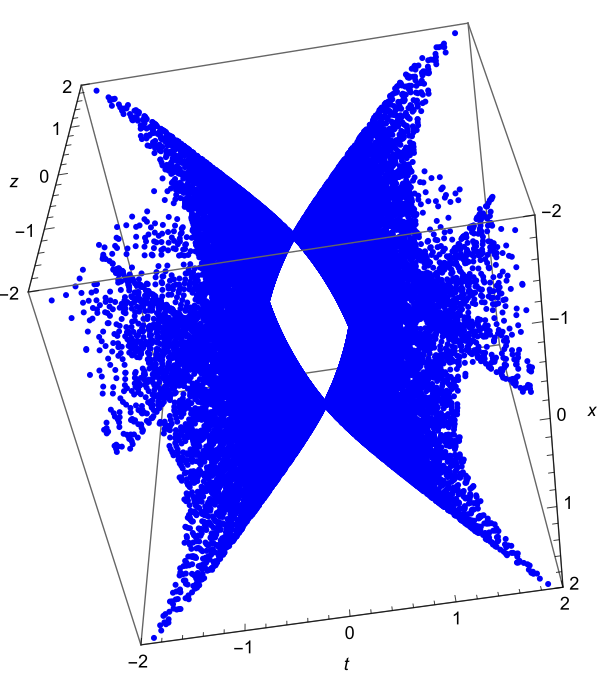}}
\end{minipage}
\caption{Geometric representation of $|\theta\rangle_-$ for the cut-off ${\cal{N}}= 200$,  $\varphi_A = 0.4$, $\varphi_B=0.6$ and, respectively, $r_A=1.2$, $r_B=1.3$ (a), $r_A=r_B=1$ (b).}
\label{geo1}
\end{figure}

Exemplary numerical solutions are depicted in figure \ref{geo1}. 
Here we restricted ourselves to antisymmetric states $| \theta\rangle_-$. For symmetric states, $| \theta\rangle_+$, one finds similar results. We also ignored the effective contributions, treating all points as equivalent. Figure \ref{geo1} (a) and figure \ref{geo1} (b) represent two different sets of parameters labeling the states, respectively for $r_i\neq 1$ and $r_i=1$. Consider figure \ref{geo1} (a) first. As we see, there are two perpendicular lines at the boundary\footnote{By the ''boundary'' we understand the conformal boundary of the anti-de Sitter space corresponding to $z=0$.} $z=0$, surrounded by some ''fluctuations'' in the bulk, composed of  points in space which do not form any specific structure. The lines coincide with world lines of two massless particles propagating in opposite directions at the boundary, and crossing at a single point $(0,0,0)$. What is interesting here is that the points extracted from the partial states are grouped mostly along these world lines. As such, they form distinct, quasi-continuous structures.

\begin{figure}[!htb]
\begin{minipage}{1\linewidth}
  \centering
  \subfloat[$t=0.1$]{\includegraphics[width=0.25\textwidth]{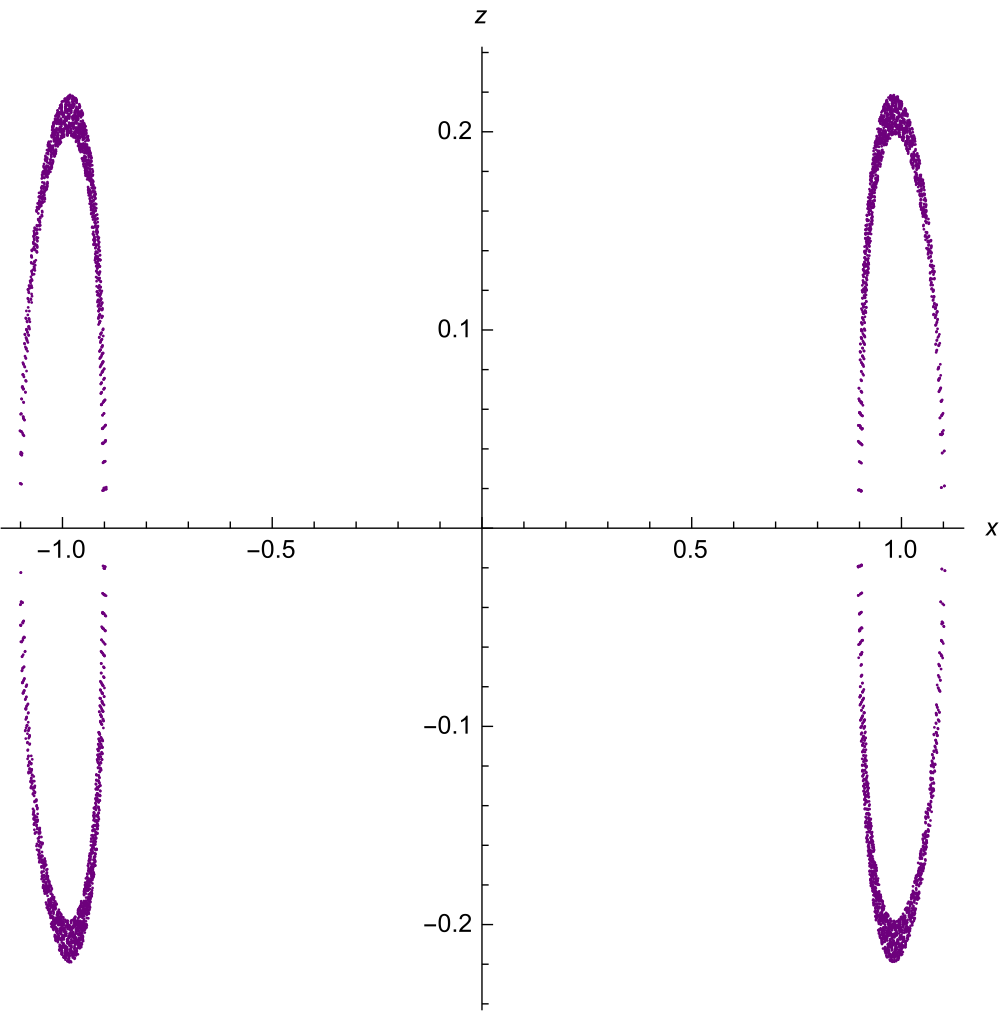}} 
  \hfill
  \subfloat[$t=0.45$]{\includegraphics[width=0.25\textwidth]{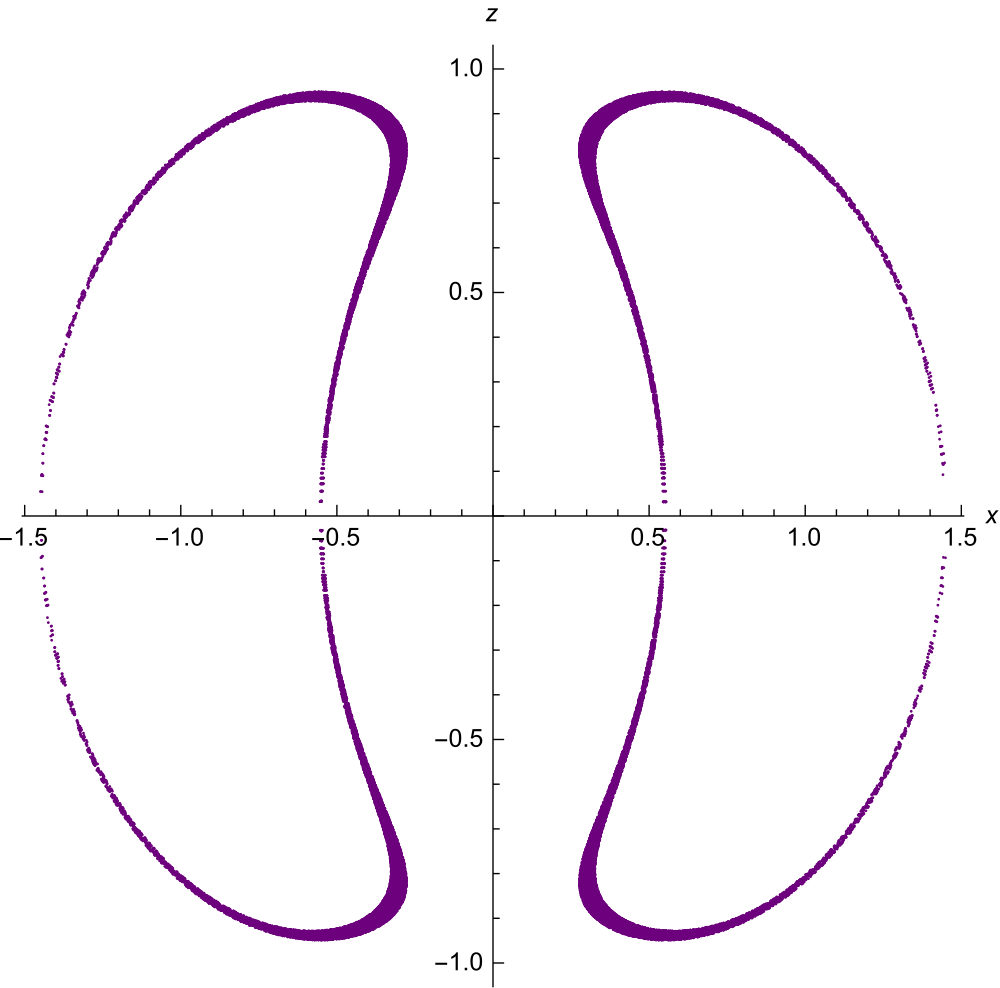}}
 \hfill
  \subfloat[$t=0.47$]{\includegraphics[width=0.25\textwidth]{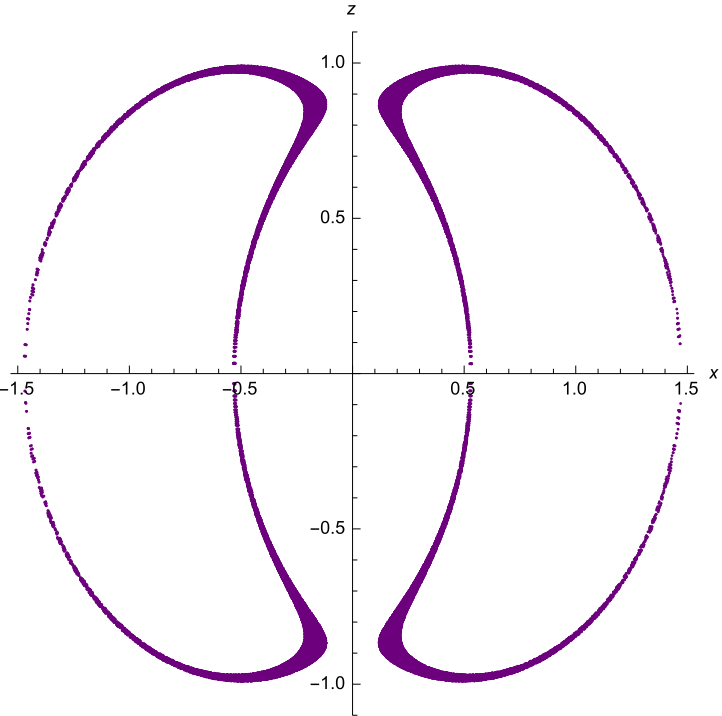}}
\end{minipage}\par\medskip
\begin{minipage}{1\linewidth}
  \centering
  \subfloat[$t=0.48$]{\includegraphics[width=0.25\textwidth]{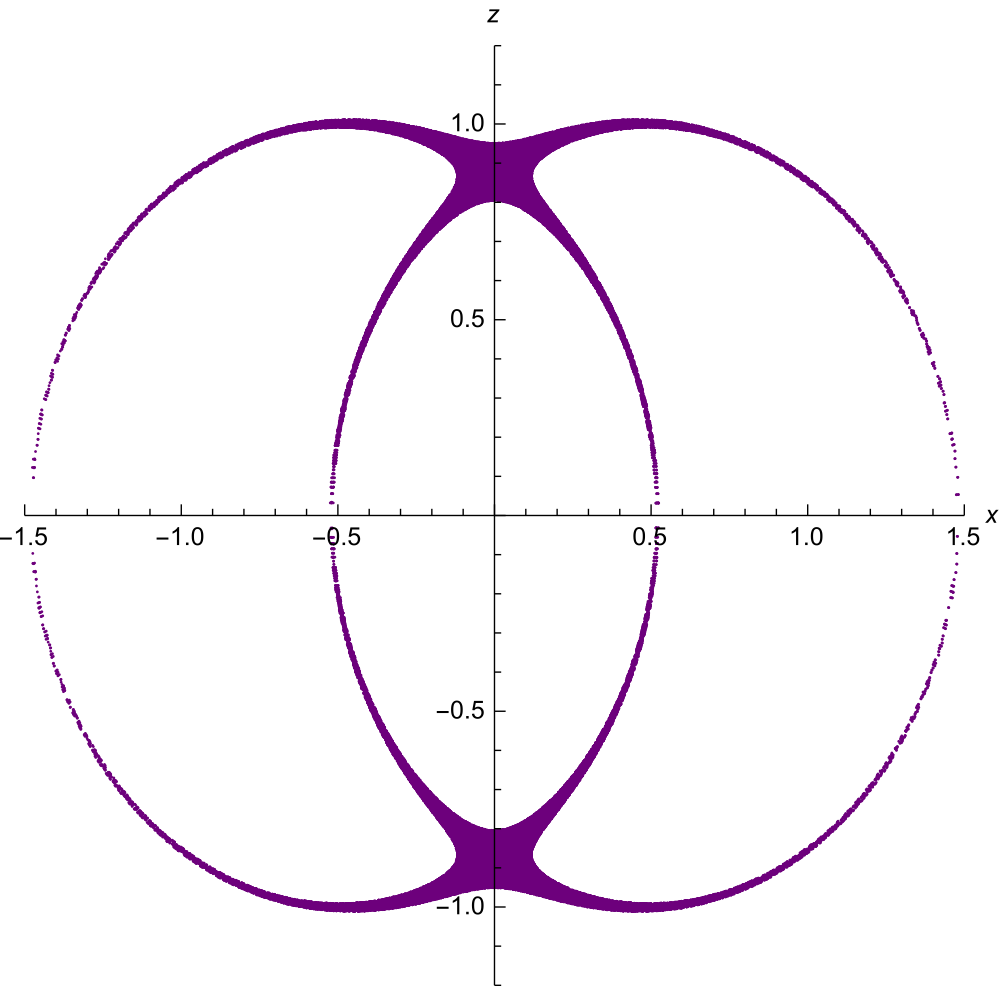}}
  \hfill
  \subfloat[$t=0.49$]{\includegraphics[width=0.25\textwidth]{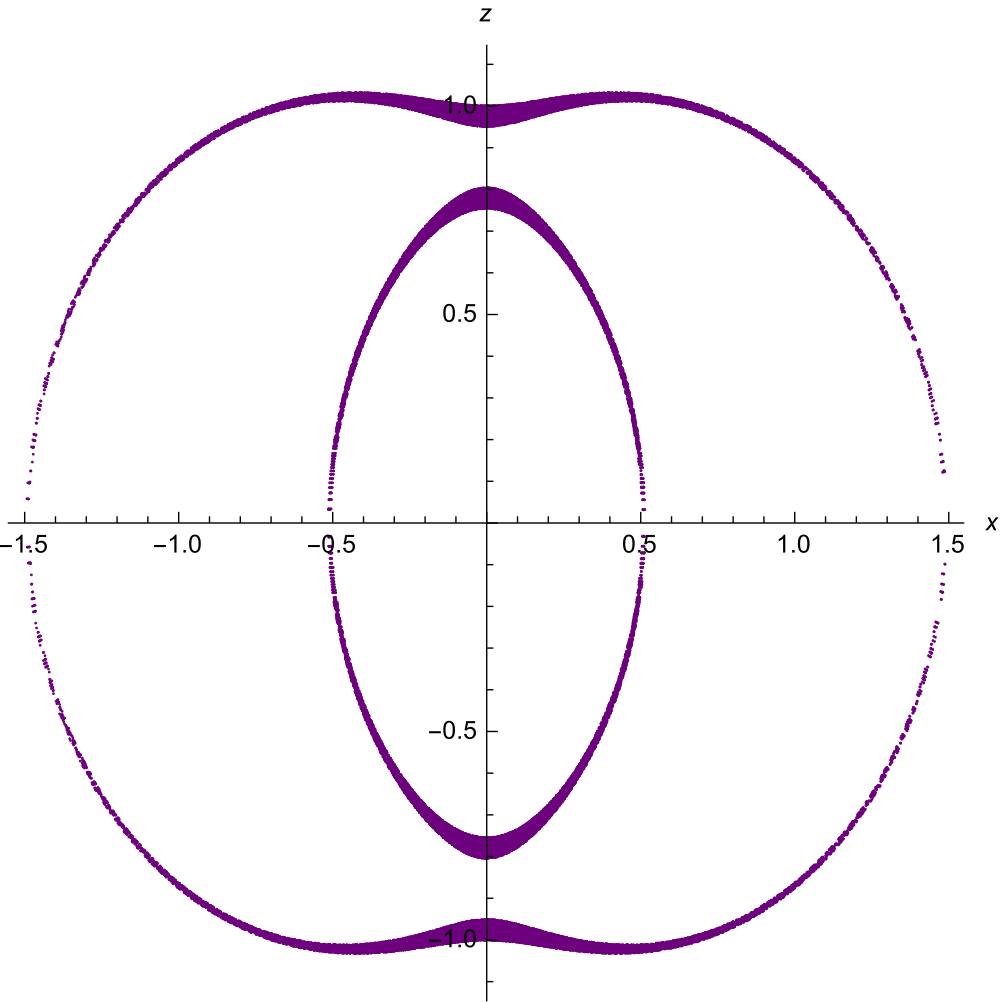}}
  \hfill
  \subfloat[$t=0.8$]{\includegraphics[width=0.25\textwidth]{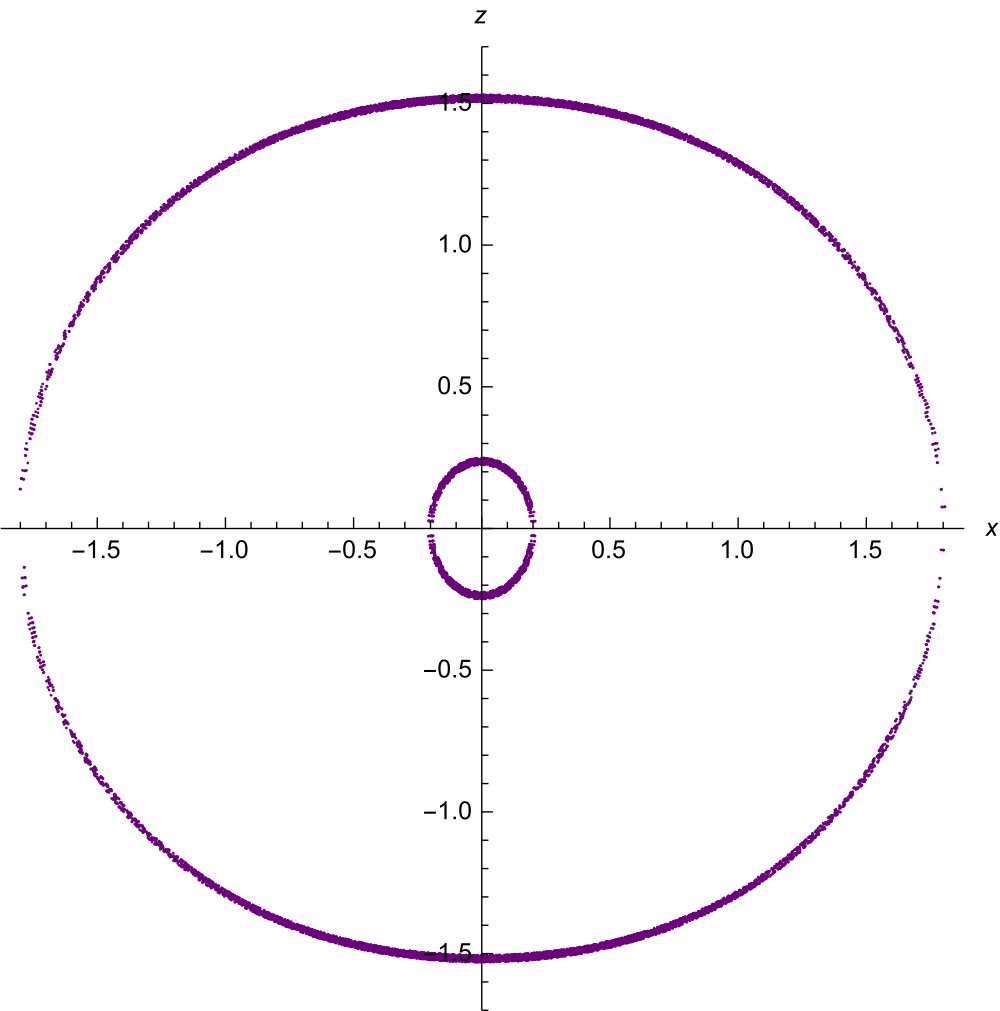}}
\end{minipage}\par\medskip
\begin{minipage}{1\linewidth}
  \centering
  \subfloat[$t=1$]{\includegraphics[width=0.25\textwidth]{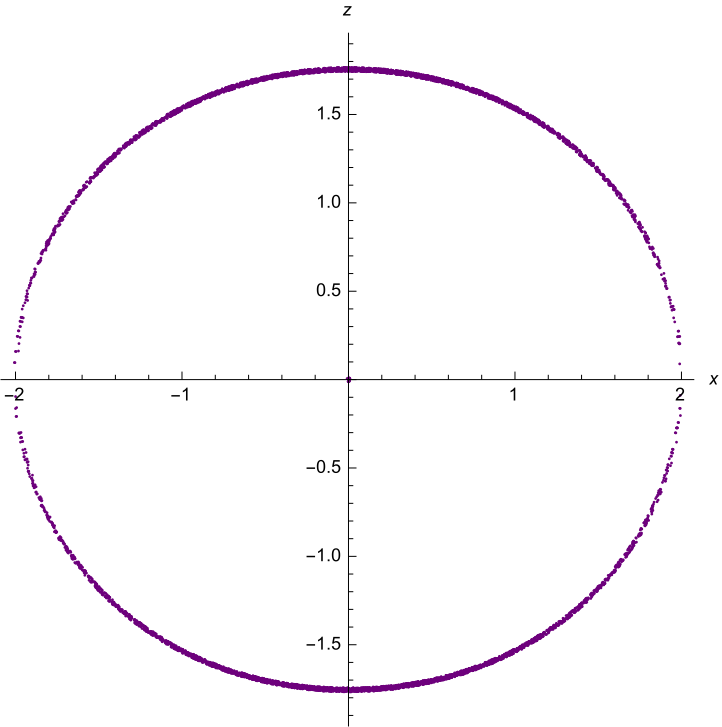}} 
  \hfill
  \subfloat[$t=1.5$]{\includegraphics[width=0.25\textwidth]{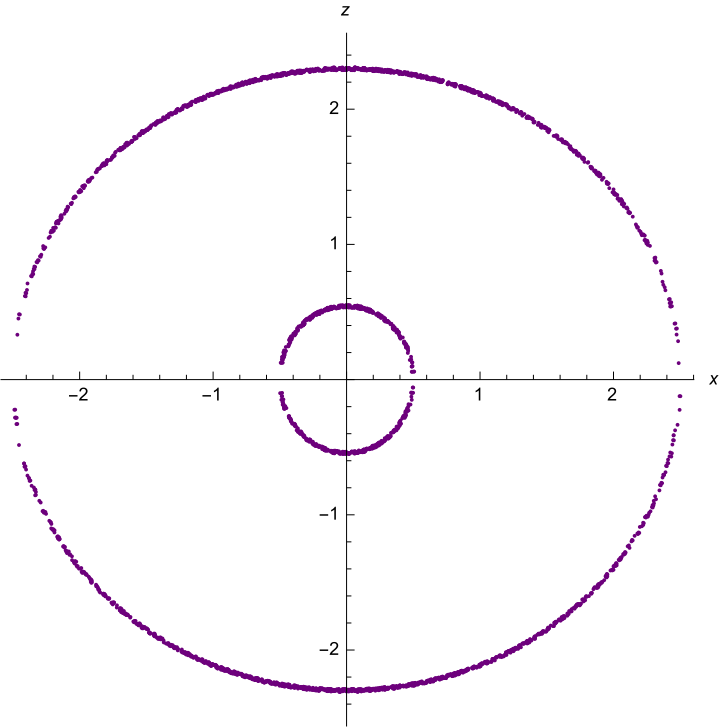}}
 \hfill
  \subfloat[$t=4$]{\includegraphics[width=0.25\textwidth]{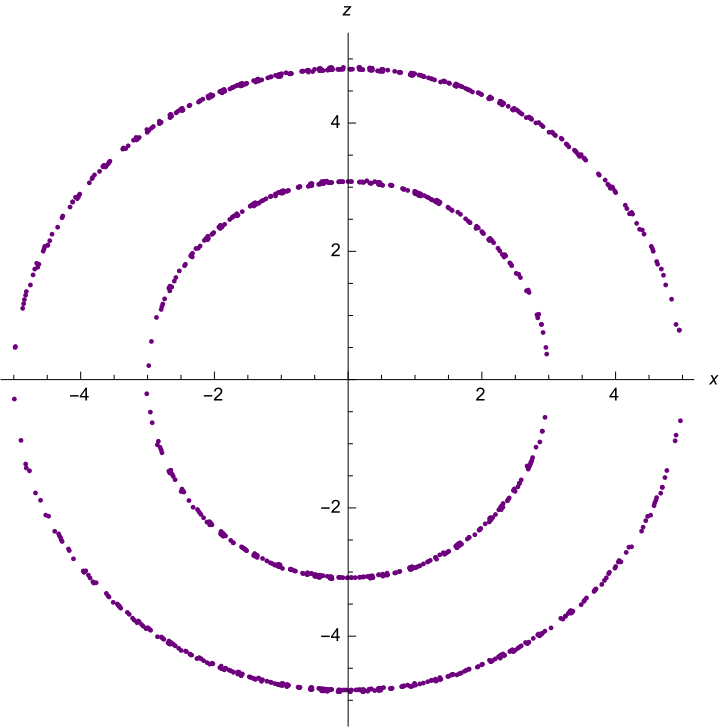}}
\end{minipage}\par\medskip
 \caption{Cross sections $t=const$ of the graph presented in figure \ref{geo1} (b). To improve the resolution the cut-off was increased to ${\cal{N}}=1500$. The hypersurfaces of constant time were chosen as rectangular regions of non zero thickness, typically of order $\delta t \simeq 0.01$.}
\label{geo2}
\end{figure}

A different situation is presented in figure \ref{geo1} (b). Here, instead of world lines at the boundary, we observe a non-trivial structure in the bulk. However, the three-dimensional plot does not reveal much of the details. A much better illustration is to consider cross sections of constant time. Due to the only finite number of discrete points in space, the latter should be replaced  by a cuboid  of non-zero thickness $\delta t$. This way any fixed moment in time will be associated with some additional uncertainty of order of $\delta t$. The results are presented in figure \ref{geo2}.
As we see, the structure is symmetric under parity transformations $x \rightarrow -x$ and $z \rightarrow -z$. Although it is not presented, the same holds for time reversal $t \rightarrow -t$. Note that the cross sections resemble two-dimensional closed strings. Contrary to the expectations, the ''strings'' shrink down and elongate with time. The most prominent illustration of changing in size is depicted in figure \!\ref{geo2} (g). Here one of the two strings shrank down to a point. However, a more detailed analysis with a higher resolution (greater $\cal{N}$) shows that this is actually not a point, but another closed string. Still, it is more likely there is a moment in time the string become arbitrarily small, expanding shortly thereafter (see figure \ref{geo2} (h)). The expansions and contractions can be regarded as an effect related to the choice of coordinates. In particular, it is easily to check that at least for a perfectly spherical closed string in $AdS_3$ centered at $(x,z)=(0,0)$, the invariant length is zero and remains unchanged during the evolution. Another observation is that the point-like configuration is meaningless because of zero effective contribution. Referring to numerical calculations, it can be found that $|c_{nm}|^2 \rightarrow 0$ once $t$ or $z$ are becoming small. In particular, this is why in figure \ref{geo2} we skipped the cross section corresponding to $t=0$.

Note the strings ''interact'' by splitting and joining together, as depicted in figure \ref{geo2} (b)-(f). This behaviour is an interesting analogue of interactions in string theory. In fact, this is why we identified the cross sections of constant time with the closed strings. Since we do not consider string theory, perhaps it would be better to call them ''loops''. Whatever they are called, they are loop-like objects resembling a worldsheet of interacting closed strings.

The later stages of evolution are presented in figure \ref{geo2} (h)-(i). As time passes, the strings become increasingly spherical, approaching close to each other. Although it is not presented, more detailed numerical simulations show that they merge eventually. The main difficulty in providing the corresponding graphical illustration lies in the fact that this requires enlarging drastically the cut-off $\cal{N}$. From a certain moment in time, it would be hard to recognize strings, loops or structures of any kind.

Having said this, it is time to summarize the results we have achieved so far. First of all, discussing geometric interpretation of the maximally entangled states \eqref{zBell} we restricted ourselves to the two fixed sets of parameters $\theta^a_i$. However, the results turn out to be generic. In particular, it can be shown that depending on the values of $r_i$, one identifies structures of two types:  word lines of massless particles at the boundary and worldsheets of closed string in the bulk. The latter correspond to $r_i=1$, while the former are specific for $r_i \neq 1$. The phases $\varphi_i$ play also an important role, making the bulk or boundary structures more or less visible. In the most extreme case corresponding to $\varphi_i=0$ and $r_i=1$, the whole emergent space is reduced just to two points $(t,x,z)=(0,0,\pm 1)$. A number of simulations suggest there is a general rule, stating that once a non-trivial structure arises, it takes a form of one of the two types presented above. This concerns both antisymmetric and symmetric states \eqref{zBell}.

Secondly, the results can be further modified allowing various redefinitions of the initial map \eqref{map0}. As we recall, the last specifies coordinates in $\mathbb{R}^{2,2}$. There is an additional ambiguity related to the choice of the map. For instance, one can consider the following modification of eqs. \eqref{map0}
\be
\label{modi}
X^{1}_{\pm nm}:=\frac{1}{\sqrt{2}}(\tilde{\delta}^\pm_{nm}+\tilde{\gamma}^\pm_{nm}), \quad X^{2}_{\pm nm}:=\frac{1}{\sqrt{2}}( \tilde{\zeta}^\pm_{nm} + \tilde{\xi}^\pm_{nm}).
\ee
Despite the structure of the scalar product \eqref{scalar2} remains the same, this may affect graphical illustration of the entangled states \eqref{zBell}. Analysis of several numbers of various different maps leads to the conclusion that modifications of the type \eqref{modi} may or may not change structures presented in figure \!\ref{geo1} and figure \!\ref{geo2}. However, the changes turn to be minuscule from the topological point of view. For instance, this consists in shifting the point of intersection of world lines at the boundary or in the lack of some stages of evolution of closed strings in the bulk. 

Thirdly, there is an interesting analogy with the AdS/CFT correspondence. Depending on the values of parameters $\theta^a_i$ (and so $r_i$, $\varphi_i$), we obtained two different geometric representations of maximally entangled states \eqref{zBell}. Note that the spectrum of the Hamiltonian \eqref{hamone} is independent of the choice of parameters. The last two facts, if combined together, suggest some connection between the ''bulk'' and the ''boundary''. This concerns solely our geometric interpretation of the vacuum  and has nothing to do with the true duality between different theories, predicted by the AdS/CFT correspondence. However, the geometric illustration remains very intuitive.

Fourthly, the structures presented either in fig \ref{geo1} or \ref{geo2} result in plotting points specific the entangled states in Poincare coordinates. On the other hand, anti-de Sitter space was discovered studying the structure of the scalar product \eqref{vform1}, independently of any particular choice of coordinates. What makes the choice important is the fact the states \eqref{zBell} result only in a discrete number of points in $AdS_3$, covering only a non-trivial regions in space. These were examined adopting Poincare coordinate system. From this perspective an interesting question is the role of coordinates. In particular, one could ask if it is possible to find a solution with the Einstein-Rosen bridge. If so, the corresponding coordinates would be better from perspective of ER=EPR conjecture. Possibly, a good option would searching for a method of installing the coordinates depending on the state in consideration, or finding manifestly coordinate-independent way of visualization. These would give a better insight into the subject. Still, Poincare coordinates can be justified as a simple choice revealing a distinct kind of bulk-boundary duality: massless particles at the boundary and closed string-like objects in the bulk.  

We close the section coming back to the structure presented in figure \ref{geo1} (a). As we recall, this was interpreted as representing  pairs of particles living at the conformal boundary of the anti-de Sitter space. We also pointed out the analogy to AdS/CFT. In fact, in the context of AdS/CFT correspondence the structure can be identified with a pair of entangled particles in the dual CFT \cite{Jensen:2013ora,Sonner:2013mba,Chernicoff:2013iga}. In particular, \cite{Chernicoff:2013iga} shows that there is an essential connection between quantum entanglement and Hawking radiation of the pair. The entangled particles are connected in the bulk  by an open string with a non-trivial worldsheet causal structure of the form of a two-dimensional Einstein-Rosen bridge. The last holds true even for pure anti-de Sitter space. At this point it is worth mentioning our results: the Hawking effect caused by breaking entanglement in the vacuum (the emission process \eqref{emix}), the presence of anti-de Sitter space and  identification of pairs of particles at the boundary. It would be interesting to ask if the casual structure can be examined by more standard methods, finding correlators of the oscillators \cite{deBoer:2017xdk,Chaudhuri:2018ihk,Banerjee:2018twd}. Most likely, this would require  generalized extended operators introduced in appendix \ref{genext} and introducing more direct equivalents of standard quantum fields. Despite technically involved, the construction can reveal even deeper connection between quantum entanglement and spacetime geometry, predicted by ER=EPR conjecture.

\section{Summary}
\label{sum}

The main goal of this paper was to construct a framework illustrating how, in principle, quantum entanglement can be connected with spacetime. This was done considering the concept of external operators in the extended Hilbert space. Starting with definition \ref{defi1}, we reformulated the quantum harmonic oscillator identifying the basic elements: creation and annihilation operators, the Hilbert space and the Hamiltonian. The results were obtained in a new manner and, in particular, independently of the classical system. For instance, we found creation and annihilation operators searching for the simplest combinations of external operators defining normalized states, the Hilbert space emerges as nothing but a positive definite sector related with the algebra, the algebra comes from basic commutation relations of the external operators, etc. All of these facts justify our initial postulates, making them an interesting starting point for a more complicated analysis. 

In the next step we have shown that the two-dimensional quantum harmonic oscillator can be reformulated in such a way the ground states is maximally entangled and takes the form of one of the two Bell states $|\theta\rangle_\pm$.  We observed that breaking the entanglement in the vacuum leads to emission of a quantum of energy. This can be used to provide  a simple toy model, illustrating the Hawking radiation process as a process of breaking the entanglement in the vacuum. In this model the  maximally entangled ground state of the two-dimensional oscillator plays a role of the vacuum state of relativistic QFT in a fixed gravitational background of a black hole.

A unique feature of the construction is that all states of quantum harmonic oscillator have a unique expansion in the extended space. Starting with the two-dimensional harmonic oscillators constructed from either symmetric $|\theta\rangle_+$ or antisymmetric $|\theta\rangle_-$ maximally entangled ground states, we examined expansions of both the vacuum states $|\theta\rangle_\pm$, finding them to be interpreted geometrically. The interpretation bases on three observations. The first is that  each of the maximally entangled state $|\theta\rangle_\pm$ itself  is a superposition of infinite number of other orthonormal, maximally entangled states, the so-called partial states $|\theta_{nm} \rangle_\pm$. They form a convenient basis in the extended space. The second is the structure of the partial states: we showed they can be associated with unit vectors in $\mathbb{R}^{2,2}$, and so points in $AdS_3$. Finally, combining the results we concluded that the states $|\theta\rangle_\pm$ can be also interpreted geometrically, identifying them  with discrete regions in $AdS_3$. 

Plotting points corresponding to the partial states on a single graph we uncover the structure of the regions specific the entangled states  $|\theta\rangle_\pm$. Depending on the values of internal parameters $\theta^a_i$ labeling the states, this reveals structures of two types. The first is a pair of massless particles at the boundary. The second resemble world sheets of interacting closed strings in the bulk $AdS_3$. Since the choice of parameters is meaningless from perspective of the oscillator (it does not affect the spectrum of the Hamiltonian) the twofold geometrical interpretation suggests a kind of a bulk-boundary duality. The analogy is can be even stronger recalling the results found in the context of the AdS/CFT correspondence. I particular, as predicted in \cite{Chernicoff:2013iga}, a pair of entangled particles in CFT can be viewed as connected in the bulk by an open string with a worldsheet causal structure in the form of a two-dimensional Einstein-Rosen bridge. This is the case even if the bulk has the form of pure anti-de Sitter space.

Summing up, the main message of this paper is finding a tool allowing to interpret geometrically a very simple class of maximally entangled states. This was done examining geometrical interpretation of the vacuum of the two-dimensional quantum harmonic oscillator and identifying a tree-dimensional anti-de Sitter space in result. The latter appears as an emergent concept: being absent from the beginning, the spacetime was ''extracted'' from the form of entangled vacuum. Anti-de Sitter space was obtained in an entirely new manner and, what is important, aside from the AdS/CFT correspondence and any gravitational theory.  This suggests an even stronger connection between quantum mechanics and gravity, indicated in \cite{Susskind:2017ney}. However, the two-dimensional oscillator presented in the paper is not the goal by its own. This is only a simple example illustrating how the connection with geometry can be established within the formalism. The key point here is that entanglement is essential for geometric interpretation. As predicted by the ER=EPR conjecture, the state need to be maximally entangled in order to be interpreted geometrically.

The construction can be generalized in many different ways. For instance, one can use generalized operators presented in appendix \ref{genext} to construct more realistic states. In particular,  one can ask about possible geometric interpretation of the vacuum QFT or examine causal structure of space by computing two and four-point correlation functions \cite{deBoer:2017xdk,Chaudhuri:2018ihk,Banerjee:2018twd}. All of this can shed more light on connection with geometry and proving that quantum entanglement might be crucial in understanding the nature of gravity.

\bigskip

\appendix
\section{Generalized operators}
\label{genext}

We now generalize the results of section \!\ref{extended} assuming the external operators are labeled  by discrete $i$, as well as continuous $x \in \mathbb{R}^d$ labels, i.e. $(e_i,\partial_i) \rightarrow (e_i(x),\partial_i(x))$. The required  modification of the definition \ref{defi1} reads
\begin{align}
 [e_i(x),e_j(y)]=[\partial_i(x),\partial_j(y)]:=0,
\\[1ex]
 \langle \varphi | e_{i_1}^{n_1}(x_1)...e_{i_k}^{n_k}(x_k)  | \phi \rangle = \langle \varphi | \partial_{i_1}^{n_1}(x_1)...\partial_{i_k}^{n_k}(x_k)  | \phi \rangle := 0,
\\[1ex]
 e_i^\dagger(x) := e_i(x), \quad \partial_i^\dagger(x) := -\partial_i(x),
\\[1ex]
 \partial_i(x) e_j(y) | \phi \rangle := \delta_{ij}\delta^d(x-y)| \phi \rangle.
\end{align}
Additionally we require
\begin{align}
\nonumber
\partial_i(x) e_{j_1}(y_1)...e_{j_n}(y_n)|\phi \rangle &= \delta_{i  j_1} \delta^d(x-y_1) e_{j_2}(y_2)...e_{j_n}(y_n)|\phi \rangle +...
\\[1ex]
\label{gchain1}
&+ \delta_{i j_n} \delta^d(x-y_n) e_{j_1}(y_1)...e_{j_{n-1}}(y_{n-1})|\phi \rangle,
\\[1ex]
\nonumber
e_i(x) \partial_{j_1}(y_1)...\partial_{j_n}(y_n)|\phi \rangle &= -\delta_{i j_1} \delta^d(x-y_1) \partial_{j_2}(y_2)...\partial_{j_n}(y_n)|\phi \rangle +...
\\[1ex] 
\label{gchain2}
&-\delta_{i j_n}\delta^d(x-y_n) \partial_{j_1}(y_1)...\partial_{j_{n-1}}(y_{n-1})|\phi \rangle.
\end{align}
The operators satisfy the following commutation relation
\begin{equation}
\label{genxpalgebra2}
[\partial_j(y),e_i(x) ] = \delta_{ij} \delta^d(x-y)\hat{\eta}.
\end{equation}
A linear combination of $e_i(x)$ and $\partial_i(x)$  specifies two normalized orthogonal states of the form $\alpha_i(x)|\phi\rangle$, $\alpha_i^\dagger(x)|\phi\rangle$, where
\begin{align}
\label{alphafinal2}
\alpha_i(x) := (\theta^1_i + i \theta^2_i) e_i(x) + \left(\frac{1-2 \theta^2_i \theta^3_i}{2 \theta^1_i}+ i \theta^3_i  \right) \partial_i(x).
\end{align}
Indeed, one checks
\begin{align}
\nonumber
&\langle \phi| \alpha_i(x)  \, \alpha_j(y) |\phi \rangle = \langle \phi| \alpha_i^\dagger(x)  \, \alpha_j^\dagger(y)  |\phi \rangle = 0,
\\[1ex]
&\langle \phi| \alpha_i(x) \, \alpha_j^\dagger(y) |\phi \rangle = - \langle \phi| \alpha_j^\dagger(y)  \, \alpha_i(x)  |\phi \rangle = \delta_{i j} \delta^d(x-y) \| \phi \|.
\end{align}
Similarly
\begin{align}
\nonumber
&[\alpha_i(x),\alpha_j^\dagger(y)] = i \delta_{i j} \delta^d (x-y) \hat{\eta},
\\[1ex]
\label{alg2b}
&[\alpha_i(x),\alpha_j(y)] = [\alpha_i^\dagger(x),\alpha_j^\dagger(y)]=0.
\end{align}
Note that now $\partial_i(x)$ and $e_i(x)$ are analogue of functional derivatives. Again, $e_i(x)$ ''differentiate'' with the extra minus sign.

\section{Functional coefficients}
\label{coef1}

The functional coefficients in the scalar product \eqref{vform1} read
\begin{align}
\nonumber
\gamma^-_{nm} &= b_A b_B \frac{2^{n+m+\frac{1}{2}}  n! m!}{(2n+1)!(2m+1)!} \cos\Big[\frac{\pi}{4}(1+(-1)^{n+m})-n \varphi_A-m \varphi_B \Big] r_A^n r_B^m,
\\[1ex]
\nonumber
\delta^-_{nm} &= b_A b_B \frac{2^{n+m+\frac{1}{2}}  n! m!}{(2n+1)!(2m+1)!} \cos\Big[\frac{\pi}{4}(1+(-1)^{n+m})-(n+1) \varphi_A-(m+1) \varphi_B \Big] r_A^{-n-1} r_B^{-m-1},
\\[1ex]
\nonumber
\xi^-_{nm} &= b_A b_B \frac{2^{n+m+\frac{1}{2}}  n! m!}{(2n+1)!(2m+1)!} \cos\Big[\frac{\pi}{4}(1-(-1)^{n+m})-(n+1) \varphi_A+m \varphi_B \Big] r_A^{-n-1} r_B^m,
\\[1ex]
\zeta^-_{nm} &= b_A b_B \frac{2^{n+m+\frac{1}{2}}  n! m!}{(2n+1)!(2m+1)!} \cos\Big[\frac{\pi}{4}(1-(-1)^{n+m})-n \varphi_A+(m+1) \varphi_B \Big] r_A^n r_B^{-m-1},
\end{align}

\begin{align}
\nonumber
\gamma^+_{nm} &= b_A b_B \frac{2^{n+m+\frac{1}{2}}  n! m!}{(2n+1)!(2m+1)!} \cos\Big[\frac{\pi}{4}(1-(-1)^{n+m})-n \varphi_A-m \varphi_B \Big] r_A^n r_B^m,
\\[1ex]
\nonumber
\delta^+_{nm} &= b_A b_B \frac{2^{n+m+\frac{1}{2}}  n! m!}{(2n+1)!(2m+1)!} \cos\Big[\frac{\pi}{4}(1-(-1)^{n+m})-(n+1) \varphi_A-(m+1) \varphi_B \Big] r_A^{-n-1} r_B^{-m-1},
\\[1ex]
\nonumber
\xi^+_{nm} &= b_A b_B \frac{2^{n+m+\frac{1}{2}}  n! m!}{(2n+1)!(2m+1)!} \sin\Big[\frac{\pi}{4}(1-(-1)^{n+m})-(n+1) \varphi_A+m \varphi_B \Big] r_A^{-n-1} r_B^m,
\\[1ex]
\zeta^+_{nm} &= b_A b_B \frac{2^{n+m+\frac{1}{2}}  n! m!}{(2n+1)!(2m+1)!} \sin\Big[\frac{\pi}{4}(1-(-1)^{n+m})-n \varphi_A+(m+1) \varphi_B \Big] r_A^n r_B^{-m-1},
\end{align}
where $b_{A/B}$ are given by \eqref{bi}. For the product \eqref{scalar2} the corresponding coefficients read
\begin{align}
\nonumber
\tilde{\gamma}^-_{nm} &= \frac{\sin\Big[\frac{\pi}{4}(1-(-1)^{n+m})+ n \varphi_A+ m \varphi_B \Big]}{ [1+\cos((4n+2)\varphi_A)]^\frac{1}{4} [1+\cos((4m+2)\varphi_B)]^\frac{1}{4} } r_A^{n+\frac{1}{2}} r_B^{m+\frac{1}{2}},
\\[1ex]
\nonumber
\tilde{\delta}^-_{nm} &= -\frac{\cos\Big[\frac{\pi}{4}(1+(-1)^{n+m})-(n+1) \varphi_A-(m+1) \varphi_B \Big]}{ [1+\cos((4n+2)\varphi_A)]^\frac{1}{4} [1+\cos((4m+2)\varphi_B)]^\frac{1}{4} } r_A^{-n-\frac{1}{2}} r_B^{-m-\frac{1}{2}},
\\[1ex]
\nonumber
\tilde{\xi}^-_{nm} &= \frac{\cos\Big[\frac{\pi}{4}(1-(-1)^{n+m})-(n+1) \varphi_A+ m \varphi_B \Big]}{ [1+\cos((4n+2)\varphi_A)]^\frac{1}{4} [1+\cos((4m+2)\varphi_B)]^\frac{1}{4} } r_A^{-n-\frac{1}{2}} r_B^{m+\frac{1}{2}},
\\[1ex]
\tilde{\zeta}^-_{nm} &= \frac{\cos\Big[\frac{\pi}{4}(1-(-1)^{n+m})-n \varphi_A+ (m+1) \varphi_B \Big]}{ [1+\cos((4n+2)\varphi_A)]^\frac{1}{4} [1+\cos((4m+2)\varphi_B)]^\frac{1}{4} } r_A^{n+\frac{1}{2}} r_B^{-m-\frac{1}{2}},
\end{align}

\begin{align}
\nonumber
\tilde{\gamma}^+_{nm} &= \frac{\sin\Big[\frac{\pi}{4}(1+(-1)^{n+m})+ n \varphi_A+ m \varphi_B \Big]}{ [1+\cos((4n+2)\varphi_A)]^\frac{1}{4} [1+\cos((4m+2)\varphi_B)]^\frac{1}{4} } r_A^{n+\frac{1}{2}} r_B^{m+\frac{1}{2}},
\\[1ex]
\nonumber
\tilde{\delta}^+_{nm} &= \frac{\sin\Big[\frac{\pi}{4}(1+(-1)^{n+m})+(n+1) \varphi_A+(m+1) \varphi_B \Big]}{ [1+\cos((4n+2)\varphi_A)]^\frac{1}{4} [1+\cos((4m+2)\varphi_B)]^\frac{1}{4} } r_A^{-n-\frac{1}{2}} r_B^{-m-\frac{1}{2}},
\\[1ex]
\nonumber
\tilde{\xi}^+_{nm} &= \frac{\sin\Big[\frac{\pi}{4}(1-(-1)^{n+m})-(n+1) \varphi_A+ m \varphi_B \Big]}{ [1+\cos((4n+2)\varphi_A)]^\frac{1}{4} [1+\cos((4m+2)\varphi_B)]^\frac{1}{4} } r_A^{-n-\frac{1}{2}} r_B^{m+\frac{1}{2}},
\\[1ex]
\tilde{\zeta}^+_{nm} &= -\frac{\sin\Big[\frac{\pi}{4}(1-(-1)^{n+m})-n \varphi_A+ (m+1) \varphi_B \Big]}{ [1+\cos((4n+2)\varphi_A)]^\frac{1}{4} [1+\cos((4m+2)\varphi_B)]^\frac{1}{4} } r_A^{n+\frac{1}{2}} r_B^{-m-\frac{1}{2}}.
\end{align}

\end{document}